\documentclass[12pt,nohyper]{JHEP3} 
\usepackage{epsfig}
\title{Minimal $Z^\prime$ models: \\ present bounds and early LHC reach} 
\author{ Ennio Salvioni \\
Dipartimento di Fisica, Universit\`a di Padova and INFN, \\
Sezione di Padova, Via Marzolo 8, I-35131 Padova, Italy \\
E-mail: \email{ennio.salvioni@pd.infn.it}} 
\author{ Giovanni Villadoro \\
Theory Group, Physics Department, CERN \\
CH--1211 Geneva 23, Switzerland \\
E-mail: \email{giovanni.villadoro@cern.ch}} 
\author{ Fabio Zwirner \\
Dipartimento di Fisica, Universit\`a di Padova and INFN, \\
Sezione di Padova, Via Marzolo 8, I-35131 Padova, Italy \\
E-mail: \email{fabio.zwirner@pd.infn.it}}
\preprint{CERN-PH-TH/2009-160 \\ 
DFPD-09/TH/17} 
\abstract{
We consider `minimal' $Z^\prime$ models, whose phenomenology is controlled by only three parameters beyond the Standard Model ones: the $Z^\prime$ mass and two effective coupling constants. They encompass many popular models motivated by grand unification, as well as many arising in other theoretical contexts. This parameterization takes also into account both mass and kinetic mixing effects, which we show to be sizable in some cases. After discussing the interplay between the bounds from electroweak precision tests and recent direct searches at the Tevatron, we extend our analysis to estimate the early LHC discovery potential. We consider a center-of-mass energy from 7 towards 10~TeV and an integrated luminosity from 50 to several hundred pb$^{-1}$, taking all existing bounds into account. We find that the LHC will start exploring virgin land in parameter space for $M_{Z^\prime}$ around 700~GeV, with lower masses still excluded by the Tevatron and higher masses still excluded by electroweak precision tests. Increasing the energy up to 10~TeV, the LHC will start probing a wider range of $Z^\prime$ masses and couplings, although several hundred pb$^{-1}$ will be needed to explore the regions of couplings favored by grand unification and to overcome the Tevatron bounds in the mass region around 250~GeV.
}
\begin{document} 

\section{Introduction and summary} 
\label{intro}

Among the extensions of the Standard Model (SM) at the TeV scale, those with an additional $U(1)$ factor in the gauge group, associated with a heavy neutral gauge boson $Z^\prime$, have often been considered in direct and indirect searches for new physics, and in the studies of possible early discoveries at the LHC  (for recent reviews and references, see e.g. \cite{reviews}). While not prescribed by compelling theoretical or phenomenological arguments, these extensions naturally arise from Grand Unified Theories (GUTs) based on groups of rank larger than four and from higher-dimensional constructions such as string compactifications. $Z'$ bosons also appear in little Higgs models, composite Higgs models, technicolor models and other more or less plausible scenarios for physics at the Fermi scale.

Many varieties of $Z^\prime$ models have been considered over the years \cite{reviews}. In the following, we will concentrate on a class of {\em minimal}  models, previously discussed in \cite{adh}, that stands out for its simplicity and for the small number of additional free parameters with respect to the SM ones. Nevertheless, this class is sufficiently rich and motivated to emerge as a natural benchmark for comparing direct and indirect signals in different experimental contexts, in particular for organizing experimental searches at present hadron colliders such as the Tevatron and the LHC.

By \emph{minimal} $Z'$ models we mean the most economical U(1) extensions of the SM that do not spoil renormalizability. Making reference to the SM particle content, our minimality requirements can be summarized as follows: no exotic vectors, apart from a single $Z^\prime$ associated with a $U(1)$ factor in the gauge group, commuting with $G_{SM} = SU(3)_C \times SU(2)_L \times U(1)_Y$; no exotic fermions, apart from one right-handed neutrino, singlet under $G_{SM}$, for each of the three SM families, and family-independent but generically non-vanishing $U(1)$ charges; no exotic scalars beyond the SM Higgs field, meaning that the extra $U(1)$ factor in $G$ is either broken explicitly (which is still compatible with renormalizability), or that the possible non-SM scalars from the extended Higgs sector can be neglected in the first $Z^\prime$ `discovery studies', because they are sufficiently heavy and/or sufficiently decoupled from the SM fields, including the SM Higgs.

The introduction of right-handed neutrinos makes it possible to generate a realistic pattern of neutrino masses (Dirac and Majorana) via renormalizable interactions. Similarly, it is possible to cancel all gauge and gravitational anomalies without introducing exotic fermions (as is instead required in some $E_6$ models) and/or non-renormalizable anomaly-canceling terms (as in some string-inspired models), as long as the generator of the additional $U(1)$ factor is a linear combination of the weak hypercharge $Y$ and of baryon-minus-lepton number, $B-L$. Also, our minimal class of models interpolates continuously among several discrete examples already considered in the literature: the `pure $B-L$' model, the `$\chi$' model arising from $SO(10)$ unification, left-right symmetric models, etc.  Minimal $Z^\prime$ models can be extended to include supersymmetry, but the price to pay is the introduction of many free parameters associated with the supersymmetric particle spectrum: for this reason we will restrict ourselves to the minimal case, commenting on the possible effects of light supersymmetric particles when relevant. 

As we shall see, the simplicity of the parameterization allows us to perform a relatively model-independent study, which does not focus on a fixed type or size of $Z'$ couplings and automatically takes into account mixing effects, both in the mass terms and in the kinetic terms, such as those generated in the evolution under the renormalization group equations (RGE). We find that the latter effects are indeed quite important, even for relatively weakly coupled $Z'$.

With the parameterization above, we perform an updated analysis of the present indirect bounds coming from electroweak precision tests (EWPT), including LEP2 and other 
experiments at low energy, and of the recent limits from direct searches at the Tevatron, taking into account the effects of mixing when necessary. Contrary to the common lore, we find that models with no $Z-Z'$ mixing, such as pure $B-L$ models, are as constrained as other models; actually, they are even more constrained than those with a partial cancellation of some suitable effective charge, such as the $\chi$ model. 

Our goal is to apply our parameterization to study the impact of the present bounds on the discovery reach of the LHC, especially in the early phase when energy and luminosity will be limited. We find that the first virgin land in parameter space that will become accessible to the LHC will correspond to relatively weakly coupled and light $Z'$, with masses of 600-800 GeV, while $Z^\prime$ bosons of the kind favored by GUTs, which are forced to be heavier by the present bounds, will require some more energy and luminosity to become accessible. We will show that indeed the unexplored region of parameter space---masses and types of $Z'$ couplings---that will become accessible at each energy and luminosity, especially during the first LHC runs, depends non-trivially on the present bounds. This makes the use of a model-independent parameterization, such as the one suggested in this work, a valuable tool to systematically organize the $Z'$ searches.

Our paper is organized as follows. In Section~\ref{theory}, we elaborate on some results of \cite{adh} and illustrate how, in our minimal class of models, the main  $Z^\prime$ properties and its couplings to the SM states can be very simply described in terms of just three new parameters beyond the SM ones: the $Z^\prime$ mass, $M_{Z^\prime}$, and two effective coupling constants, $g_Y$ and $g_{BL}$, associated to the $Y$ and $B-L$ currents, respectively\footnote{The reader should not confuse $g_Y$ with the SM $U(1)_Y$ gauge coupling constant, which we denote by $g'$, with $g$ denoting the $SU(2)_L$ gauge coupling constant.}. Such parameterization automatically takes into account all mixing effects in the gauge boson sector, both in the kinetic terms and in the mass matrix. In contrast with GUT-inspired parameterizations, often used for interpreting experimental searches, our parameterization also allows for relatively weak $Z^\prime$ couplings to the SM fermions: such a situation is disfavored by conventional GUTs but could arise, for example, in anisotropic string compactifications. To give a flavor of the generic GUT constraints, we solve the one-loop RGE for the above coupling constants, both in the non-supersymmetric and in the supersymmetric case, showing that the large hierarchy between the GUT scale and the weak scale generically induces sizable kinetic mixing effects from the running, even for moderate values of the coupling constants. We then identify both the region of parameter space generically favored by GUTs and some points in parameter space corresponding to `true' GUT-models, comparing them with some conventional benchmark points often used in the literature. In Section~\ref{bounds}, we review the present bounds on the $Z^\prime$ parameter space. We start by discussing the impact of EWPT, including those from LEP2 and atomic parity violation (APV), applying the methods and the results of \cite{CCMS}. In addition, we discuss the impact of some later reanalysis of APV \cite{apvth}, to conclude that the electroweak fit of \cite{CCMS} still gives the most stringent constraints. Then we discuss the bounds coming from direct searches at the Tevatron, including the most recent data from CDF \cite{CDFepem,CDFmumu} and D0 \cite{D0epem}. We confirm that, as for the models considered in previous studies \cite{bprs,carena,contino}, EWPT are more stringent than Tevatron direct searches for relatively heavy and relatively strongly coupled $Z^\prime$, such as the ones predicted by GUTs. On the other hand, Tevatron searches are more stringent than EWPT for relatively light and relatively weakly coupled $Z^\prime$, disfavored by GUTs but potentially permitted by other models. As a representative example, we comment on the possible excess in the CDF $e^+e^-$ sample \cite{CDFepem} at an invariant mass near 240 GeV: the size of the effect would correspond to a rather weakly coupled $Z^\prime$, disfavored by GUTs but still allowed by EWPT. We conclude in Section~\ref{LHC} with a study of the prospects for the first LHC run(s). The main question we address is what energy and luminosity are needed to explore virgin land in parameter space, and how much of such accessible new territory is compatible with conventional GUTs. We show that, with the foreseen schedule for the first year \cite{earlylhc}  of the LHC (first 50-100 pb$^{-1}$ at $\sqrt{s}=7$~TeV, then up to 200$\div$300 pb$^{-1}$ at $\sqrt{s} \le 10$~TeV), the first region in parameter space to be explored will correspond to moderately light and weakly coupled $Z'$, weighing around 600-800~GeV, and with a small window of allowed couplings. To open up considerably the region of parameter space accessible for discoveries, in particular the one relevant to GUT models, at least $O(1)$~fb$^{-1}$ of integrated luminosity should be collected. In summary, in the very first phase of the LHC the interplay among center-of-mass energy,  integrated luminosity and previous direct and indirect bounds will be quite subtle, and it will be important to focus the analysis onto the most promising regions of parameter space, possibly combining different channels and experiments from the very beginning. 

\section{Theory} 
\label{theory}

\subsection{Parameterization} 
\label{param}

As discussed in the Introduction, we will consider extensions of the SM where $G_{SM}$ is extended by a single additional non-anomalous family-independent $U(1)$ factor, in the presence of three full SM fermion families, including right-handed neutrinos. As for the Higgs sector of the theory, we will assume the existence of the SM Higgs field but avoid as much as possible any specific assumption on the symmetry breaking mechanism for the additional $U(1)$.

As previously discussed in \cite{adh}, it is not restrictive to parameterize masses, kinetic mixing \cite{kinmix} and interactions with fermions for our extended neutral electroweak sector by means of the following effective Lagrangian:
\begin{equation} 
\label{LAB}
\mathcal{L} = - \frac{1}{4} \, h_{AB} \, F_{\mu\nu}^{A} \, F^{B\,\mu\nu} + \frac{1}{2} \, M^{2}_{AB} \, A^{A\mu} \, A_{\mu}^{B} + A_{\mu}^{A} \, J^{\mu}_{A} + \ldots \, , 
\end{equation}
where $A,B=T_{3L},Y,B-L$. It is also well known that, by appropriate field redefinitions, we can go to a field basis where kinetic terms are canonical and masses are diagonal:
\begin{equation}
\label{Lij}
\mathcal{L} = - \frac{1}{4} \, \, F_{\mu\nu}^{i} \, F^{i\,\mu\nu} + \frac{1}{2} \, M^{2}_{i} 
A^{i \mu} \, A_{\mu}^{i} + A_{\mu}^{i} \, J^{\mu}_{i} + \ldots \, , 
\end{equation}
where $i=\gamma , Z , Z'$. In the above equation, $M_\gamma$, $M_Z$ and 
$M_{Z^\prime}$ are the mass eigenvalues. The currents $J^\mu_i$ ($i= \gamma, Z, Z^\prime$) are those coupled to the gauge boson mass eigenstates. For example,
\begin{equation}
J^\mu_\gamma = e \, \sum_f Q(f) \, \overline{f} \gamma^\mu f 
\end{equation}
is the electromagnetic current, where $f$ runs over the different chiral projections of the SM fermions, $Q(f)=T_{3L}(f)+Y(f)$ is their electric charge, and the contributions from the scalar sector have been omitted. Similarly, we can write
\begin{equation}
J_Z^\mu  =  \cos \theta^\prime \, J_{Z^0}^\mu -  \sin \theta^\prime \, J_{Z^{\prime \, 0}}^\mu  \, , 
\qquad
J_{Z^\prime}^\mu  =  \sin \theta^\prime \, J_{Z^0}^\mu +  \cos \theta^\prime \, J_{Z^{\prime \, 0}}^\mu  \, , 
\end{equation}
where
\begin{equation}
J_{Z^0}^\mu = g_Z  \, \sum_f \left[ T_{3L}(f) - \sin^2 \theta_W \, Q(f) \right]  \, \overline{f} \gamma^\mu f  \, , 
\qquad
\left( g_Z = \sqrt{g^2+g^{\prime \, 2}} \right) \, , 
\label{zcurr}
\end{equation}
is the SM expression for the current coupled to the SM $Z^0$ (we recall that, in the presence of mixing, $Z^0$ does not coincide with the mass eigenstate $Z$), and
\begin{eqnarray}
 J_{Z^{\prime \, 0}}^\mu &=& \sum_f \left[ g_Y \, Y(f) + g_{BL} \, (B-L) (f) \right] \, \overline{f} \gamma^\mu f \nonumber \\
  &=& \sum_f g_{Z} \, Q_{Z'}(f) \, \overline{f} \gamma^\mu f   \, .
\label{zpcurr}
\end{eqnarray}
Again, possible contributions to the currents from the scalar sector have been omitted.
We collected in Tab.~\ref{charges} the charges of the SM fermions needed for evaluating the currents of eqs.~(\ref{zcurr}) and (\ref{zpcurr}). For definiteness, we chose a purely left-handed basis for the fermion fields, so that, omitting family indices, $f = u , d , u^c , d^c , \nu , e , \nu^c , e^c $. In expressing the charges $Q_{Z'}$,  we found it convenient to make reference to the ratios 
\begin{equation}
\label{ratios}
\widetilde{g}_Y = \frac{g_Y}{g_Z} \, , 
\qquad
\widetilde{g}_{BL} = \frac{g_{BL}}{g_Z} \, . 
\end{equation}
\TABLE[t]{
\begin{tabular} {|c | c | c | c | c | c | c |  }
\hline
& $(u,d)$ & $u^c$ & $d^c$ & $(\nu,e)$  & $\nu^c$ & $e^c$ \\
\hline
$T_{3L}$ & $(+\frac12,-\frac12)$  & $0$ & $0$ & $(+\frac12,-\frac12)$ & $0$ & $0$ \\   
\hline
$Y$ & $+ \frac16$  & $- \frac23$ & $+\frac13$ & $- \frac12$  & $0$ & $+1$ \\   
\hline
$B-L$ & $+ \frac13$  & $- \frac13$ & $-\frac13$ & $-1$ & $+1$ & $+1$ \\
\hline
$Q_{Z'}$ & $\frac16 \widetilde g_Y+ \frac13\widetilde g_{BL}$  & 
$-\frac23 \widetilde g_Y- \frac13\widetilde g_{BL}$ & $\frac13 \widetilde g_Y- \frac13\widetilde g_{BL}$ & 
$-\frac12 \widetilde g_Y-\widetilde g_{BL}$ & 
$\widetilde g_{BL}$ & $\widetilde g_Y+\widetilde g_{BL}$ \\
\hline
\end{tabular}
\caption{The charges of left-handed fermions controlling the electroweak neutral currents.}
\label{charges}
}

The parameterization above automatically contains and extends specific models often considered in the literature, such as $Z_{B-L}$, $Z_\chi$, and $Z_{3R}$ models, whose couplings simply read, in our notation:
\begin{equation} 
\label{eq:chargemodels}
\begin{array}{r|ccc}
		& Z_{B-L} & Z_{\chi} & Z_{3R} \\ \hline
g_Y		& 0	& \phantom{-}-\frac{2}{\sqrt{10}}g_{Z'}\phantom{-} & \phantom{-}-g_{Z'}\phantom{-} \\
g_{B-L}	&\phantom{-}\sqrt{\frac38}g_{Z'} \phantom{-}& \phantom{-}\frac{5}{2\sqrt{10}}g_{Z'}&\phantom{-}\frac12 g_{Z'}
\end{array} \ ,
\end{equation}
where $g_{Z'}$ is usually fixed to a `GUT-inspired' value $g_{Z'}=\sqrt{5/3} \, g'$.

Since the SM Higgs doublet $H$ has\footnote{This property is shared by the MSSM Higgs doublets $H_1$ and $H_2$.} vanishing $B-L$, and, as discussed in \cite{adh}, it is not restrictive to take the Higgs fields that break $B-L$ (if any) to have vanishing $Y$, we can express the $Z-Z^\prime$ mixing angle $\theta^\prime$ in terms of $g_Y$ and $M_{Z^\prime}$,
\begin{equation}
\tan \theta^\prime = - \widetilde{g}_Y \, \frac{M_{Z^0}^2}{M_{Z^\prime}^2 - M_{Z^0}^2} \, ,
\end{equation}
where
\begin{equation}
M_{Z^0}^2 = \frac{g_Z^2 \, v^2}{4}
\end{equation}
is the SM expression for the $Z^0$ mass. The same remains true if we assume that there is an explicit (or St\"uckelberg-like) diagonal mass term for the $Z^{0 \, \prime}$, without introducing an additional complex Higgs field for breaking $B-L$.

Notice that the mixing angle is completely determined by the mass and the couplings of the $Z'$. In particular, it is always non-vanishing whenever $\widetilde g_Y\neq0$ (i.e. for models different from pure $B-L$), because in these cases gauge invariance of the Yukawa terms forces the SM Higgs to be charged under the extra $U(1)$, thus producing a $Z-Z'$ mixing.

We can then study the $Z^\prime$ phenomenology in terms of three unknown parameters: the $Z^\prime$ mass $M_{Z^\prime}$ and the two coupling constants $(g_Y,g_{BL})$ or, equivalently, $(\widetilde g_Y,\widetilde g_{BL})$.

We will not consider possible additional parameters of the enlarged Higgs sector and the right-handed neutrino masses, because, as will be discussed later, these parameters will play a relatively minor r\^ole in the following. To be definite, we will assume that there are three mostly left-handed neutrinos lighter than $O(1)$~eV and three mostly right-handed neutrinos heavier than $M_{Z^\prime}/2$, as in the see-saw mechanism, and that the physical components of the Higgs fields whose VEVs break $B-L$ (if any) have negligible mixing with the SM Higgs and masses larger than $M_{Z^\prime}$. 

\subsection{Constraints from grand unification} 
\label{gut}

One of the possible motivations for considering $Z^\prime$ models are GUTs, with or without supersymmetry. Through appropriate boundary conditions at the unification scale $M_U$ and RGE on the running gauge coupling constants, GUTs can constrain the range of some low-energy $Z^\prime$ parameters, such as the coupling constants $g_Y$ and $g_{BL}$. The most stringent constraints can be obtained within specific models, where the full particle spectrum is specified and threshold and higher-loop corrections can be computed. Here, instead, we would like to remain as model-independent as possible within the general class of minimal $Z^\prime$ models. To this end, we will identify a GUT-favored region in the $(\widetilde{g}_Y, \widetilde{g}_{BL})$ plane\footnote{All the results are insensitive to the transformation $(\widetilde{g}_Y, \widetilde{g}_{BL})\to (-\widetilde{g}_Y, -\widetilde{g}_{BL})$, without lack of generality we will thus consider only the upper half plane $\widetilde g_{BL}>0$.} according to the following procedure.

First, we recall that the gauge coupling constants $g^\prime$, $g_Y$, $g_{BL}$ are related to the $2 \times 2$ submatrix $h_{AB}$ with $A,B=\{Y,B-L\}$ appearing in eq.~(\ref{LAB}), via
\begin{equation} 
\label{gtohab}
h_{AB}=\left ( \begin{array}{ccc} 
\displaystyle \frac{1}{{g'}^2} &\phantom{111}& \displaystyle -\frac{g_Y}{g_{BL}}\frac{1}{{g'}^2} \\&&\\
\displaystyle -\frac{g_Y}{g_{BL}}\frac{1}{{g'}^2} && \displaystyle \frac{1}{g_{BL}^2}+\frac{g_Y^2}{g_{BL}^2}\frac{1}{{g'}^2}\end{array} \right) 
\,.
\end{equation}
The matrix $h_{AB}$ obeys simple one-loop RGEs, which can be solved analytically:
\begin{equation}
\label{habsol}
h_{AB} (M_U) = h_{AB} (M_Z) - \frac{b_{AB}}{(4 \pi)^2} \log \left(\frac{M_U}{M_Z}\right)^2 \, , 
\end{equation}
where
\begin{equation}
\label{bab}
b_{AB} = \frac{2}{3} \sum_f Q_f^A Q_f^B + \frac{1}{3} \sum_s Q_s^A Q_s^B \, ,
\end{equation}
and $s$ are the complex scalars in the theory. We collect some representative values of the $b_{AB}$ coefficients in Table~\ref{tab-bAB}, always including the contribution of the three fermion families of Table~\ref{charges}, and adding:
\begin{itemize}
\item[(i)]
only the SM Higgs field, $H \sim (1,2,+1/2,0)$;
\item[(ii)]
the SM Higgs field $H$ plus a complex SM-singlet scalar $\phi \sim (1,1,0,+2)$;
\item[(iii)]
spin-0 superpartners for all the quarks and leptons in Tab.~\ref{charges}, spin-1/2 superpartners for all the gauge bosons, and two Higgs chiral superfields $H_1 \sim (1,2,-1/2,0)$ and $H_2 \sim (1,2,+1/2,0)$, as in the MSSM;
\item[(iv)]
all the fields of the previous case, plus two extra chiral superfields $\phi_1 \sim (1,1,0,-2)$ and  $\phi_2 \sim (1,1,0,+2)$.
\end{itemize}
The quantum numbers in brackets denote, in a self-explanatory notation, the representations of $SU(3)_C$, $SU(2)_L$, $U(1)_Y$ and $U(1)_{BL}$.

\TABLE[t]{
\begin{tabular} {| c | c | c | c | c | }
\hline
\phantom{-----------------} &\phantom{---} (i)\phantom{---}   &\phantom{---}  (ii)\phantom{---}  &\phantom{--}  (iii)\phantom{--}  &\phantom{--}  (iv)\phantom{--}  \\
\hline
$b_{YY}$ & 41/6  & 41/6 & 11 & 11 \\
\hline 
$b_{Y(B-L)}$ & 16/3 & 16/3  & 8 & 8 \\   
\hline
$b_{(B-L)(B-L)}$ & 32/3 & 12  & 16 & 24 \\   
\hline
\end{tabular}
\caption{The $b_{AB}$ coefficients for the four representative cases defined in the text.}
\label{tab-bAB}
}

For choosing the boundary conditions at the GUT scale $M_U$, we normalize all $U(1)$ charges as in $SO(10)$, and we take $M_U = 10^{16} \, {\rm GeV}$ as a reference value. In typical GUTs, $M_U$  can vary within approximately two decades around such reference value, but the difference in our estimate of the GUT-favored region is of the order of other threshold effects that we reabsorbed in the wide ranges we assume below for other parameters. Then, we compute the boundary value $g^\prime (M_U)$ using the phenomenological input $g^\prime (M_Z) = e(M_Z)/\cos \theta_W (M_Z)$, with $\alpha_{em}^{-1}(M_Z) \simeq 128$ and $\sin^2 \theta_W (M_Z) \simeq 0.23$, and the SM one-loop RGE. We then allow the $Z'$ coupling at the unification scale $\alpha_U = g_U^2/(4 \pi) = g_{Z'}^2 (M_U) /(4 \pi)$, to vary within the generous bounds
\begin{equation}
\label{aurange}
\frac{1}{100} < \alpha_U < \frac{1}{20} \, . 
\end{equation}
Taking into account that the SM RGE would predict $\alpha_U \sim 1/45$, our upper and lower bounds leave a margin of more than a factor of two to account for threshold corrections, new particles at the TeV scale and other model-dependent effects. Correspondingly, we determine the GUT-favored region of the $(\widetilde{g}_Y, \widetilde{g}_{BL})$ plane by making use of the one-loop RGE of eqs.~(\ref{gtohab})-(\ref{bab}): the result is presented as the colored band in Fig.~\ref{fig-gut}.
%
\FIGURE[t]{
\includegraphics[scale=0.35]{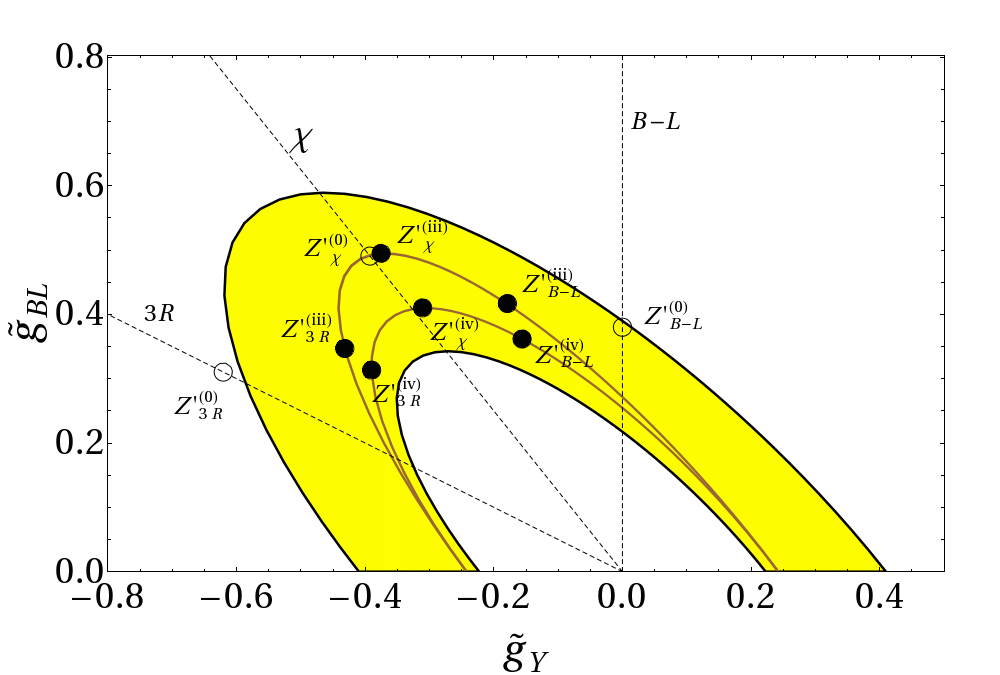} 
\label{fig-gut}
\vspace{-0.5cm}
\caption{GUT-favored region and some representative models in the $(\widetilde{g}_Y, \widetilde{g}_{BL})$ plane, see the text for details.}
}
The same figure also shows some dots that represent either some popular GUT-inspired benchmark models considered in experimental analyses (the three empty dots and the corresponding dashed lines) or specific SUSY-GUT models with an extra $U(1)$ (the three pairs of full dots). In particular, and in counter-clockwise order: the three dashed lines correspond to the three different models of eq.~(\ref{eq:chargemodels}), when $g_{Z'}$ is left free to vary; the three empty dots correspond to the GUT-inspired normalization $g_{Z'} = \sqrt{5/3}  \, g'(M_Z)$. Instead the SUSY-GUT models are derived properly, using the RGEs: they assume that the GUT group, say $SO(10)$, is broken at $M_U$ into the SM gauge group times an additional $U(1)$ factor, with charges fixed as in eq.~(\ref{eq:chargemodels}) at the GUT scale. For each of the three models (which correspond, in counter-clockwise order, to those in eq.~(\ref{eq:chargemodels})) we draw two black points, corresponding to the RGE evolutions of case (iii) (outer points) and  case (iv) (inner points). In both cases we get $M_U \sim 2 \times 10^{16} \, {\rm GeV}$ and $\alpha_U \sim 1/24$, within the bounds of eq.~(\ref{aurange}).  We do not consider non-supersymmetric GUTs among our examples, because they would require the introduction of rather ad hoc exotic fields at intermediate mass scales to match the measured gauge coupling constants at the weak scale. Notice that the introduction of the extra chiral superfields $\phi_1$ and $\phi_2$ makes the coupling constants $g_Y$ and $g_{BL}$ more `infrared free' but does not change $\alpha_U$, thus the values of $g_Y$ and $g_{BL}$ at the weak scale are smaller. 

An important point to notice is that, even if we start form a `pure $B-L$' or `pure $T_{3R}$' model at $M_U$, the mixing effects in the RGE generate, through the resummed large logarithms, sizable corrections to the effective weak-scale couplings, as can be seen from the displacement of the black dots ${Z'}_{3R/(B-L)}^{(iii)/(iv)}$ from the corresponding dashed lines in Fig.~\ref{fig-gut}. \emph{With enough running, a specific $Z'$ at the unification scale can turn into a completely different one at the weak scale!} Such effects make clear the advantage of considering the full $(\widetilde g_Y,\widetilde g_{BL})$ plane to parameterize $Z'$ searches.  Notice finally that the direction in the $(\widetilde{g}_Y, \widetilde{g}_{BL})$ plane corresponding to the $\chi$ model is quite stable under RGE, because running effects are caused by the MSSM Higgs superfields only. The RGE seem to exhibit an infrared \emph{attractor} towards the region of parameter space with $\widetilde g_Y\sim -\widetilde g_{BL}$: this fact will become even more interesting after compiling the experimental bounds. 
%
%
\section{Present bounds} 
\label{bounds}

\subsection{Electroweak precision tests} 
\label{ewpt}

The use of electroweak precision tests to put bounds on the $Z^\prime$ parameters has a long history. Pre-LEP bounds \cite{prelep} became much more stringent \cite{lep1fits} after LEP1 data at the $Z$ peak, which strongly constrain the mixing angle $\theta^\prime$. However, also higher-energy LEP2 data and APV play a very important and complementary r\^ole in constraining the $Z^\prime$ mass for given couplings, as emphasized for example in \cite{bprs,CCMS,carena,contino,erler}. In \cite{CCMS} it was shown that the bounds from EWPT can be conveniently rewritten into bounds for the nine EW pseudo-observables ($\hat S$, $\hat T$, $\hat U$, $V$, $W$, $X$,  $Y$, $\delta \epsilon_q$, $\delta C_q$). For the case of extra $Z'$ models, these pseudo-observables have simple expressions in terms of the $Z'$ couplings and mass. Once such bounds are rewritten in terms of our parameters ($M_{Z'}$, $\widetilde g_Y$, $\widetilde g_{BL}$), for each $Z'$ mass we can extract the corresponding 95\% CL exclusion regions in the $(\widetilde{g}_Y, \widetilde{g}_{BL})$ plane, as shown in Fig.~\ref{fig-ewpt}. The region allowed by EWPT is the red one enclosed by each contour. Dashed lines and empty (full) dots remind us of the GUT-favored region and of the GUT-inspired (-derived) benchmark models. We remind the reader that EWPT only constrain the ratio $g_{Z'}/M_{Z'}$: this explains why the size of the red regions in Fig.~\ref{fig-ewpt} grows linearly with the $Z'$ mass.

%
\FIGURE[t]{
\includegraphics[scale=0.35]{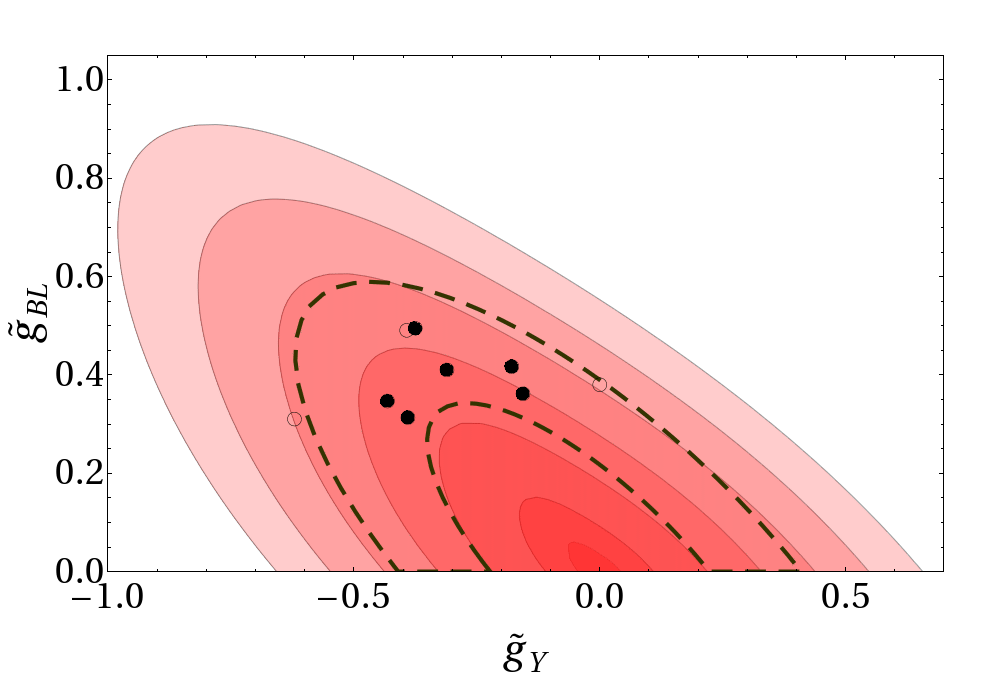} 
\label{fig-ewpt}
\caption{The regions of the $(\widetilde{g}_Y, \widetilde{g}_{BL})$ plane allowed by EWPT, at 95\% CL, for $M_{Z^\prime}=200$, $500$, $1000$, $1500$, $2000$, $2500$, $3000$~GeV (from inner to outer). The GUT-favored region is between the dashed lines.}
}

It is worth noticing that a recent theoretical re-analysis \cite{apvth} of the most precise measurements on APV \cite{apvex} was not included in the fit of ref.~\cite{CCMS}. As will be discussed later, the new bounds from APV are significantly stronger than before, but not strong enough yet to compete with the result of the global fit.

It is usually thought that pure $B-L$ models are less constrained by EWPT because of the absence of $Z-Z'$ mixing. Notice however that the region with ${\widetilde g}_Y=0$ is not particularly favored, actually the region of parameter space least constrained by EWPT is that with $\widetilde g_Y\simeq -\widetilde g_{BL}$. This feature can be understood by looking at the last row of Tab.~\ref{charges}, which shows that the $Z'$ is less coupled to matter fields, thus less constrained by LEP2 bounds (and by Tevatron bounds as well, as we will see in the next section), roughly when $\widetilde g_Y \simeq  - \widetilde g_{BL}$.

Notice also the correlation between the orientation of the GUT-favored region of Fig.~\ref{fig-gut} (between the dashed contours in Fig.~\ref{fig-ewpt}) and  the EWPT-allowed regions of Fig.~\ref{fig-ewpt}. For all values of the couplings in the GUT-favored region, and in particular for the SUSY-GUT models represented by the full dots, the lower bound on $M_{Z^\prime}$ is above 1~TeV. It is typically above $1.5\div2$~TeV for the GUT-inspired benchmark models often considered in the experimental literature. For comparison, we report in Tab.~\ref{tab-mzbounds} the bounds on the $Z'$ masses for the particular choices of the couplings corresponding to the GUT-inspired and the SUSY-GUT benchmark points\footnote{The alert reader will notice that the bounds in Tab.~\ref{tab-mzbounds} are numerically stronger than those from Fig.~\ref{fig-ewpt}: this is simply because the figure refers to a 2-parameter fit ($\widetilde g_Y,\widetilde g_{BL}$), whilst the table refers to specific models, thus to a 1-parameter ($M_Z'$) fit.} of Fig.~\ref{fig-gut}.
\TABLE[t] {
\begin{tabular}{|c||c|c|c|c|c|c|c|c|c|} 
\hline 
$\phantom{\sqrt{\frac11}}$ &${Z'}_{B-L}^{(0)}$ & ${Z'}_{B-L}^{\rm (iii)}$ &${Z'}_{B-L}^{\rm (iv)}$ 
& ${Z'}_\chi^{(0)}$   &${Z'}_\chi^{\rm (iii)}$ &${Z'}_\chi^{\rm (iv)}$
&${Z'}_{3R}^{(0)}$&${Z'}_{3R}^{\rm (iii)}$   & ${Z'}_{3R}^{\rm (iv)}$ \\ \hline \hline
$M_{Z'}$ (TeV) & 2.36 & 1.94 & 1.68 & 1.85 & 1.89 & 1.57 & 2.18 & 1.47 & 1.33 \\ \hline
\end{tabular}
\label{tab-mzbounds}
\caption{95\% CL bounds on the $Z'$ masses from EWPT, corresponding to the specific models represented by the nine points in Fig.~(\ref{fig-gut}). The ${Z'}^{(0)}$ models are those represented by empty points, while ${Z'}^{\rm(iii)}$ (${Z'}^{\rm(iv)}$) corresponds to the three external (internal) black points; see the text for details on the choice of the effective couplings.}
}

Finally, we comment on the Higgs mass dependence of the bounds above. Since the SM Higgs mass is unknown, we may worry about the stability of our fits to EWPT with respect to varying the Higgs mass. Notice, however, that the dependence of the EWPT on the Higgs mass is only logarithmic and, although $Z'$ bosons may help weakening the EWPT bounds on the Higgs mass \cite{ferroglia}, in our minimal models the preferred value for the Higgs mass is still below the LEP bound. Varying the Higgs mass within the 95\% CL limit from EWPT (which corresponds to $m_h\sim 200$~GeV)  produces only a tiny shift in the regions plotted in Fig.~\ref{fig-ewpt}. To be definite, we chose $m_h=120$~GeV as a representative value.



\subsection{Tevatron direct searches} 
\label{tevatron}

Other important bounds on $Z'$ parameters come from direct searches at hadron colliders, presently dominated by the Tevatron experiments CDF and D0. In these experiments, $Z'$ bosons of sufficiently low mass can be  produced on-shell and decay in the process $q \bar q \to Z' \to \ell^+ \ell^-$, ($\ell = e , \mu$). These two are very clean channels to look for: a peak in the invariant mass distribution of $e^+e^-$ or $\mu^+\mu^-$ pairs, with a width controlled by the experimental resolution when the intrinsic $Z'$ width is sufficiently small. The irreducible background is dominated by SM Drell-Yan (DY) $\ell^+ \ell^-$ production, which is well understood and whose control is only limited by PDF uncertainties. Other irreducible backgrounds are small and the reducible ones can be easily eliminated by generous cuts. Different decay channels into SM final states have been also experimentally investigated \cite{difcha}, for example $\tau^+ \tau^-, jet \ jet, W^+ W^-$, but for the minimal models considered here they are not competitive for exclusion or discovery: at best, they could play a r\^ole in the determination of the $Z^\prime$ couplings after a future discovery.

The procedure used to extract bounds from such processes has been extensively discussed in the literature (see for example \cite{carena,altrihad}). For a weakly coupled $Z'$, it consists in calculating the $Z'$ production cross-section  multiplied by the branching ratio into two charged leptons $\sigma(p\bar p\to Z'X) \times {\rm BR}(Z'\to \ell^+ \ell^-)$ (in our case a function of the three $Z'$ parameters $M_{Z'}$, $g_Y$, $g_{BL}$), and comparing it with the limits established by the experiments.

On the theory side, we performed the calculation at NLO in QCD (and LO for the EW part), using the NLO MSTW08 PDF sets \cite{mstw08}. In the calculation of the total width $\Gamma_{Z'}$  we included the following channels: $Z'\to f\bar f$, $W^+W^-$, and $Zh$,  where $h$ is the SM Higgs boson and $f$ are the SM fermions of Tab.~\ref{charges},  with the exception of the right-handed neutrinos, which we took to be heavier than $M_{Z'}/2$. The presence of the two last decay channels is due to $Z-Z'$ mixing and is usually neglected, however for large $Z'$ masses there is an enhancement that cancels the suppression due to the mixing \cite{zpww,zpzh}. The ratio $\Gamma_{Z^\prime} / M_{Z'}$ is pretty constant over the whole range of masses of interest, and is around 2\% for GUT-favored $Z'$ couplings, and of course smaller for more weakly coupled $Z'$. 

Notice that in the presence of extra matter fields charged under the extra $U(1)$ (light right-handed neutrinos, more light Higgses, supersymmetric partners, etc.), $\Gamma_{Z^\prime}$ would be larger, with a consequent suppression in the branching ratio to charged leptons. In this case the bounds from hadron colliders (and their ability for discovery) would be weaker (unlike those from EWPT, which are quite insensitive to these extensions of the model).

For the Tevatron experimental limit we used the most recent available results from CDF (on $Z'\to e^+e^-$~\cite{CDFepem} and $Z'\to\mu^+\mu^-$~\cite{CDFmumu}) and D0 (on $Z'\to e^+e^-$~\cite{D0epem}).  They directly provide the 95\% CL bounds on the product $\sigma(p\bar p\to Z^\prime \, X)\times BR(Z^\prime \to \ell^+\ell^-)$ based on $2.5$, $2.3$, $3.6$~fb$^{-1}$ of data with $27\div38$\%, $13\div40$\%, $17\div22$\% total acceptances respectively, with the acceptances growing from smaller to larger values of $M_{Z'}$. Notice that, although D0 data refer to a higher integrated luminosity, the acceptance is smaller, making its bounds a little weaker than those from CDF.

Since not enough information is available to us to properly combine the three sets of data, for each $Z'$ mass we took the strongest bound among the three sets. A combined analysis would be highly welcome as it would probably give stronger bounds.

When compared with the computed cross-section, the experimental limits produce, for each value of $M_{Z'}$, a 95\% CL exclusion region in the  $(\widetilde g_Y, \widetilde g_{BL})$ plane, in analogy with the EWPT case. The results are summarized in Fig.~\ref{fig-tevatron}, which shows the 95\% CL allowed regions for different $M_Z'$.
\FIGURE[t]{
\includegraphics[scale=0.35]{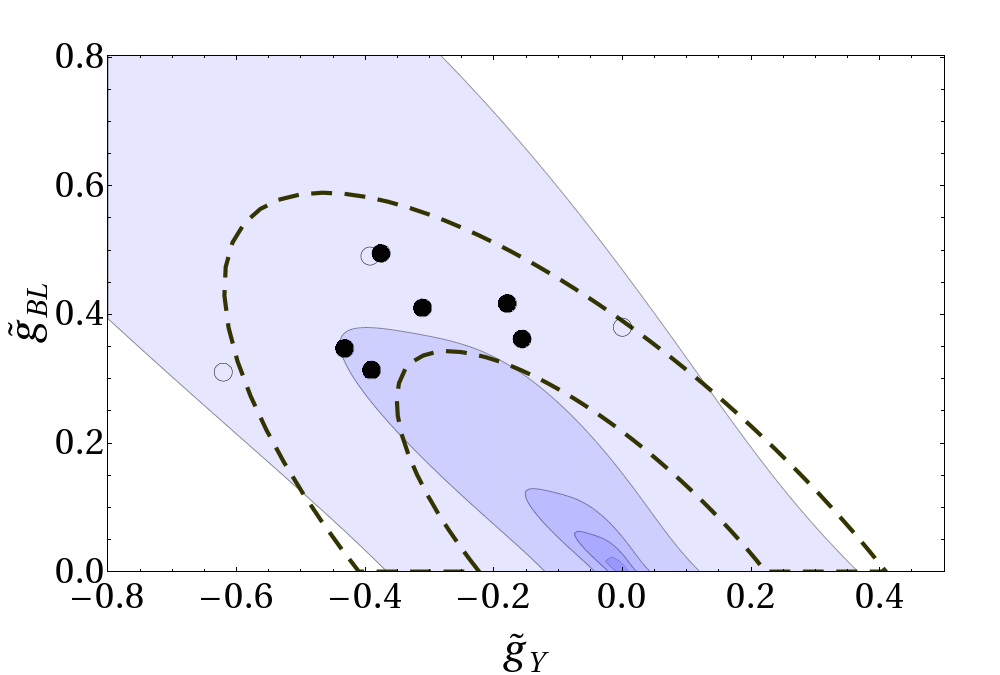} 
\label{fig-tevatron}
\caption{The regions on the $(\widetilde{g}_Y, \widetilde{g}_{BL})$ plane allowed by Tevatron direct searches at 95\% CL for $M_{Z^\prime} = 200$, $400$, $600$, $800$, $1000$~GeV (from inner to outer). The GUT-favored region is between the dashed lines.}
}
Notice that the region favored by GUT models starts becoming accessible for $M_Z' \gtrsim 700$~GeV. For masses larger than $\sim 1.2$~TeV the allowed region fills the whole plot, and the available data are no longer able to give useful constraints. Indeed, unlike the EWPT case, the allowed regions in Fig.~\ref{fig-tevatron} grow faster than linearly with $M_{Z'}$, because of the stronger suppression from the $x$-dependence of the PDF at higher energies.

\subsection{Comparison among different bounds}

We may wonder whether it makes sense to compare direct and indirect experimental bounds from different experiments, or better whether it is possible to build models that can evade indirect bounds but still be accessible to direct searches. Notice however that indirect LEP searches, low-energy APV experiments, direct and indirect searches at hadron colliders are all basically controlled by tree-level Feynman diagrams built from two basic types of elementary vertices, coupling the $Z^\prime$ to charged leptons and quarks, respectively:
\begin{center}
\includegraphics[scale=0.35]{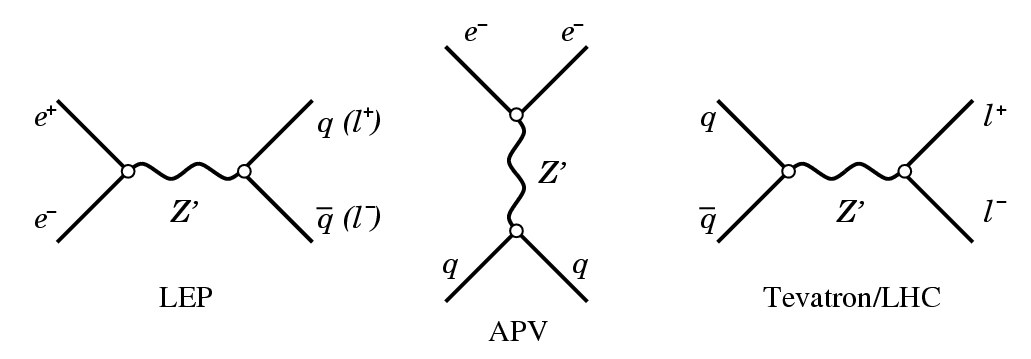}
\end{center}
Of course, LEP and APV probe off-shell $Z^\prime$ exchange, whereas the Tevatron and the LHC are sensitive to on-shell $Z^\prime$ production and decay. But the parameters involved in the relevant Feynman diagrams are the same, and it is not easy at all to invent new physics capable of evading indirect bounds from EWPT but still producing a signal in the direct searches. For this reason, the bounds from EWPT should not be neglected when analyzing the discovery reach of direct searches. On the other hand, if the branching ratio to leptons is suppressed by the presence of extra charged matter, then it may happen that indirect searches become even more powerful than the direct ones. 

The first thing we notice by comparing the plots in Figs.~\ref{fig-ewpt} and \ref{fig-tevatron} is that both EWPT and Tevatron bounds probe similar regions of the $(\widetilde g_Y,\widetilde g_{BL})$ plane. In particular, they are both less stringent in the region\footnote{Remarkably this region seems also to be an attractor of the solutions to the RGE, as commented in the previous section.} with $\widetilde g_Y\sim - \widetilde g_{BL}$, and more stringent in the orthogonal direction (although the actual shape of the exclusion region from the Tevatron is slightly different from that of EWPT). As explained in the previous section, this correlation in the couplings is due to the fact that in the region $\widetilde g_Y\sim - \widetilde g_{BL}$ there is a partial cancellation in the $Z'$-charges of the SM fermions (see Tab.~\ref{charges}). Interestingly, also the region preferred by GUT models has the same shape as those allowed by the experimental bounds: this makes the bounds on $M_{Z'}$ at fixed coupling only weakly dependent on the specific nature of the $Z'$.

Another important difference between the bounds from the Tevatron and from EWPT is that, as already noticed in some previous analyses \cite{bprs,CCMS,carena,contino}, the former give stronger constraints at smaller $M_{Z'}$ but weaker at larger $M_{Z'}$. Indeed, the possibility to produce on-shell $Z'$-bosons rewards hadron colliders in the mass region that is easily accessible to them. On the other hand, their power rapidly falls off at higher masses because of the PDF dumping at high~$x$. Already for $M_{Z'}$ around 800~GeV, the EWPT bounds start becoming more powerful than the Tevatron limits, independently of the $Z'$ couplings. In particular, all the models with GUT-favored couplings are bound much more strongly by EWPT than by direct searches, at least with the current data.

The competition between the Tevatron and EWPT is more manifest if we plot the bounds as functions of $M_{Z'}$ for a representative model. Fig.~\ref{fig-chi1} illustrates the range of $Z'$ masses where the Tevatron gives stronger or weaker constraints compared to EWPT, for a particular direction in the $(\widetilde g_Y,\widetilde g_{BL})$ plane corresponding to the so-called $Z_\chi$ models (the central dashed line in Fig.~\ref{fig-gut}).
\FIGURE[t]{
\includegraphics[scale=0.30]{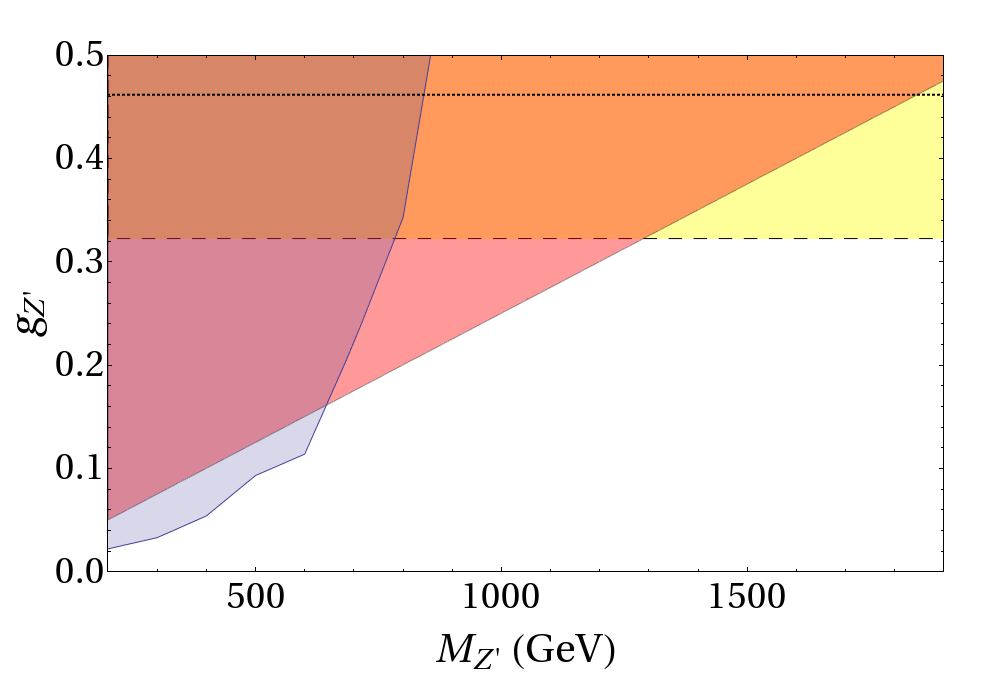} 
\label{fig-chi1}
\caption{The region of the $(M_{Z'}, g_{Z^\prime})$ plane excluded at 95\% CL by EWPT (red) and Tevatron (blue), for the $Z_\chi$ model defined in eq.~(\ref{eq:chargemodels}). The horizontal strip (yellow) recalls the GUT-preferred region and the dotted line the value $g_{Z'}=\sqrt{5/3} \, g'(M_Z)$ corresponding to the ${Z'}^{(0)}_\chi$ model.}
}
Notice that for GUT-favored couplings EWPT give much stronger bounds than the Tevatron. Similar figures were previously shown in refs.~\cite{bprs,contino}.

In Fig.~\ref{fig-apv}, as an illustration, we combine the bounds from EWPT, Tevatron direct searches and the new APV analysis of \cite{apvth}, for the two representative values $M_{Z'}=400$ and $800$~GeV. In the first case the bounds from Tevatron are the strongest for all models, while the APV bounds are the weakest. In the second case we see that EWPT start becoming stronger than those from the Tevatron in almost all parameter space: for example, Tevatron would still allow an $800$~GeV $Z'_{3R}$ with GUT-favored coupling, while EWPT would basically rule out all GUT-like $Z'$ with this mass. APV is always weaker than EWPT but starts becoming stronger than the Tevatron in the region where the latter starts performing worse, i.e. the GUT-preferred region. Notice also that APV experiments are not sensitive to certain types of $Z'$, such as pure $B-L$ that has purely vectorial couplings to fermions. However, as discussed in section~\ref{gut}, and evident in Fig.~\ref{fig-apv}, RGE effects can easily move the $Z'$ couplings away from the safe region, making such $Z'$ models, which would be otherwise safe with respect to APV bounds,  also subject to constraints.
\FIGURE[t]{
\includegraphics[scale=.2]{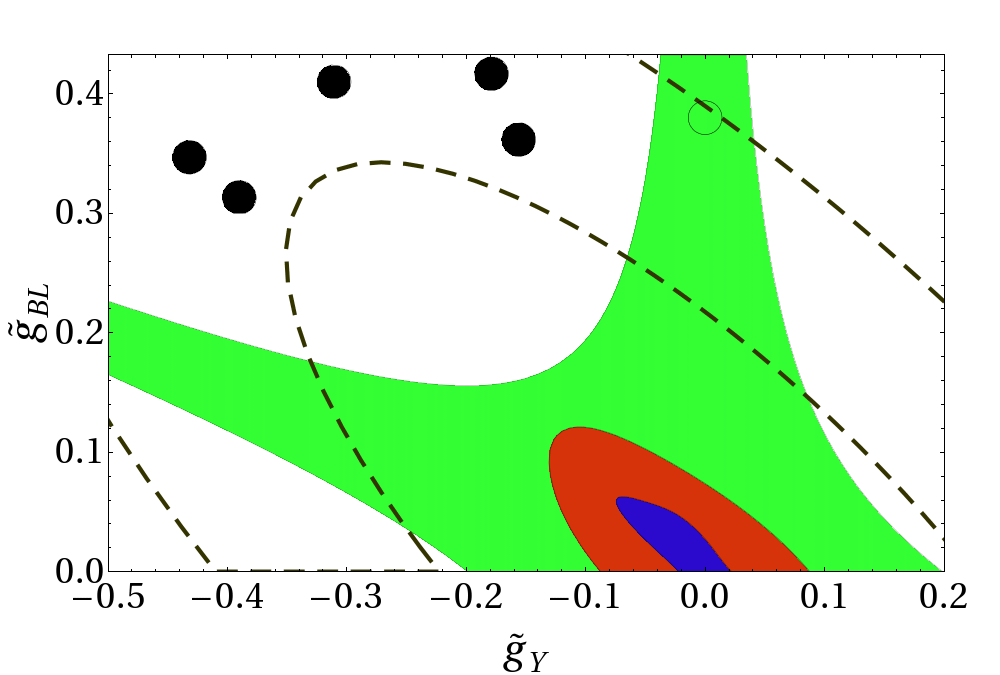} \includegraphics[scale=.2]{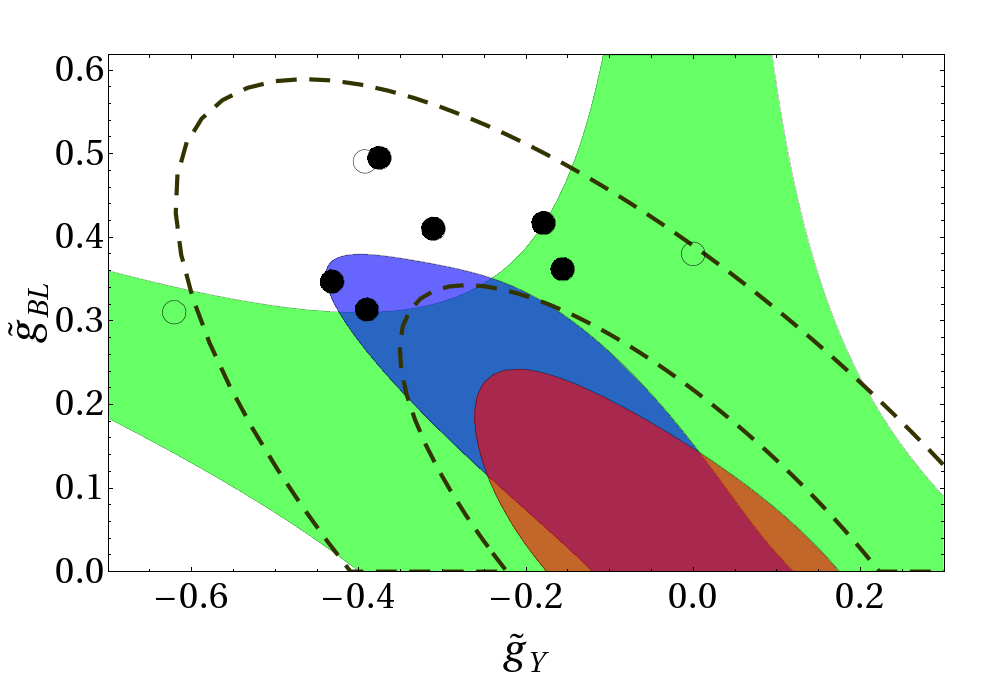} 
\label{fig-apv}
\caption{The region of the $(\widetilde{g}_Y, \widetilde{g}_{BL})$ plane allowed by EWPT (red), Tevatron (blue), and APV (green), 
for $M_{Z^\prime} = 400$ (left) and $800$~GeV (right).}
}

\subsubsection*{\it The CDF $e^+e^-$ excess}

Recently the CDF collaboration \cite{CDFepem} observed a small excess of $e^+ e^-$ events around 240 GeV, amounting to a $2.5 \, \sigma$ fluctuation from the SM background. However, CDF did not see any anomaly in the dimuon spectrum \cite{CDFmumu}. D0 data \cite{D0epem} at the moment do not seem to confirm  nor exclude such excess, also because of the smaller acceptance mentioned before. Although our minimal models would not be able to explain such effect, because of the mismatch between electron and muon spectra, we may still ask whether existing bounds from EWPT may rule out an explanation of the excess in terms of some more general $Z'$ bosons that couple non-universally to leptons, such as those discussed in \cite{nonuniv}.  Notice from Fig.~\ref{fig-chi1} that at 240~GeV Tevatron is indeed more sensitive than EWPT to neutral resonances, provided that they have a small coupling, of order $g_{Z'}\sim 0.04$. It turns out that such small coupling would be enough to explain  the Tevatron excess without contradicting EWPT\footnote{Notice that the same statement can be true also for $Z'$ models very different from those studied in this work \cite{cik}.}. We have not checked whether such scenario is compatible with other flavor non-universal low-energy constraints.  More data are anyway required to confirm the presence of a true excess and assess its possible non-standard origin. Notice instead that an analogous signal at energies higher than $\sim$700 GeV would not be compatible with EWPT, at least within the class of models considered here.


\section{Early LHC prospects} 
\label{LHC}

The questions we want to address in this section are the following. At what combined values of center-of-mass energy and integrated luminosity may we expect the LHC to start having a chance of discovering a $Z'$ (at least of the kind discussed in this work), taking into account all the experimental bounds discussed in the previous sections? What region of parameter space that has not been already ruled out could be accessible for different luminosities and energies in the first LHC runs? Considering the fact that $Z'$ signals are among the cleanest and easiest ones in the search of new physics, our analysis may also be used as a benchmark point when discussing the integrated luminosities that are worth collecting at each energy to actually probe new physics.

At the moment, the program for the first year of LHC running \cite{earlylhc} consists in a very first run at low energy ($\sqrt{s} = 7 \, {\rm TeV}$)) and low luminosity ($<100$~pb$^{-1}$), followed by an upgrade in energy ($\sqrt{s} \leq 10 \, {\rm TeV}$), with a collected luminosity up to 200$\div$300~pb$^{-1}$.

At such low\footnote{Of course, with respect to the LHC design parameters.} energies and luminosities, the constraints from Tevatron direct searches and EWPT play a crucial r\^ole in identifying the unexcluded region of parameter space that can be probed and the time scale required to have access to it.

Being the LHC a hadron collider, the region of parameter space accessible to it will be similar in shape to the corresponding one at the Tevatron. For relatively light $Z'$ ($M_{Z^\prime}  < 800$~GeV), since the strongest constraints come from Tevatron direct searches, we expect the LHC to turn into a discovery machine as soon as it becomes sensitive to regions of parameters not yet excluded by the Tevatron. However, while the higher energy is clearly a big advantage for intermediate $Z'$ masses of several hundreds GeV, for lighter masses the low luminosity may be a crucial limiting factor in the early LHC phase. On the other hand,  for heavier $Z'$ masses, such as those relevant for GUT models, generically EWPT outperform the Tevatron, and the LHC must wait for higher energies and luminosities to become sensitive.

To turn these considerations into more quantitative statements, we perform a basic analysis along the lines of the one described before for extracting the Tevatron bounds. In the present case we consider the range $\sqrt{s}=7\div10$~TeV for the $pp$ center-of-mass energy, luminosities in the range $50$~pb$^{-1}\div1$~fb$^{-1}$, and calculate the product $\sigma (pp \to Z'X) \times BR(Z'\to \ell^+\ell^-)$ for $M_{Z'}=200 \div 3000$~GeV, at the same order in perturbation theory as in the Tevatron case.
At the same level of precision, we also compute the SM Drell-Yan (DY) differential cross-section, which constitutes the main source of background.

To gain some approximate understanding of the acceptances for signal and background at different values of the invariant mass $M_{\ell^+\ell^-}$ of the $\ell^+ \ell^-$ pair, and of the possible model-dependence of the former, we performed a simple study whose results are illustrated in Fig.~\ref{fig-acc}.
\FIGURE[t]{
\includegraphics[scale=0.30]{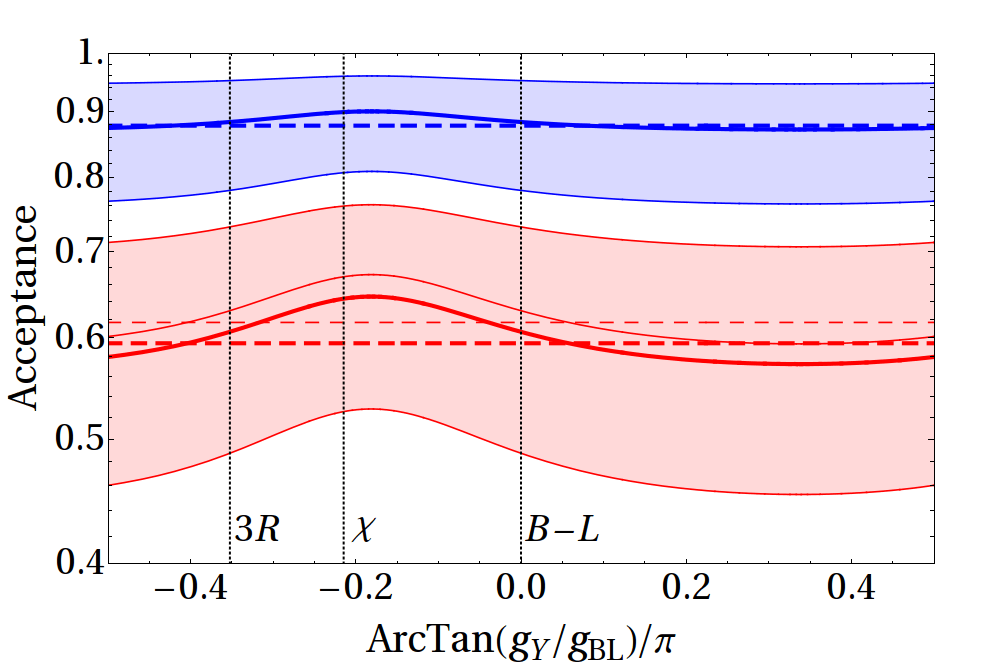} 
\label{fig-acc}
\caption{The geometrical acceptance for signal (solid lines) and SM-DY background (dashed lines), as a function of a parameter that scans over the minimal models, and for two representative values of $M_{\ell^+\ell^-}$: 200~GeV (red, lower) and 1~TeV (blue, upper). The different lines refer to the cut $|\eta|<2.5$ and  $p_{T \ell} > 20$~GeV (thin) or $p_{T \ell} > 80$~GeV (thick). The colored bands show how much the acceptance varies by changing the rapidity-cut from $|\eta|<2.1$ to $|\eta|<3.0$.}
}
For $\sqrt{s}=10 \, {\rm TeV}$, we plot the purely geometrical acceptance for signal (solid lines) and SM-DY background (dashed lines), imposing the cut $| \eta | < 2.5$. The colored bands show how much the acceptance varies if the cut is varied from $|\eta| < 2.1$ to $|\eta| < 3.0$. The quantity on the horizontal axis, $[{\rm ArcTan} \, (g_Y/g_{BL})]/\pi$, scans over the different minimal models. The upper blue lines are for $M_{\ell^+\ell^-} = 1 \, {\rm TeV}$, the lower red lines are for  $M_{\ell^+\ell^-} = 200 \, {\rm GeV}$: as expected, the acceptance for both signal and background is similar, grows with the invariant mass of the lepton-antilepton pair, and is close to 90$\%$ for $M_{\ell^+\ell^-}$ of order 1 TeV or larger. We also looked at how the acceptance depends on the cut on the lepton transverse momenta $p_{T \ell}$: the thin lines correspond to $p_{T \ell} > 20 \, {\rm GeV}$, the thick ones to $p_{T \ell} > 80 \, {\rm GeV}$. We can see from Fig.~\ref{fig-acc} that this cut is essentially included in the cut on $\eta$ for high mass values, whereas it has a small but non-negligible effect on the acceptance for low mass values. Finally, the model-dependence also decreases when moving from lower to higher masses, and is never larger than 10$\%$ even for $M_{Z^\prime} \sim 200 \, {\rm GeV}$. In view of these results, and for the purposes of the present exploratory study, we then assumed an acceptance depending only on the invariant mass, as done for example in ref.~\cite{cmsmu}. Our computed values of the acceptance are compatible with those of ref.~\cite{cmsmu}, thus we adopt their Fig.~2 for the rest of our LHC
study.  More refined studies, however, should take into account also the model-dependence of the acceptance, which may not be negligible for $Z'$ searches at relatively small masses.

To estimate the 5$\sigma$ discovery reach of the early phase of the LHC \cite{atlas,cmse,cmsmu}, we compared the events due to a generic $Z'$ signal to the events from the SM-DY background in a 3$\%$ interval around the relevant values of the dilepton invariant mass\footnote{This should be compatible with the expected energy resolution, even in this early phase, and with the fact that, for GUT-favored values of the coupling constants, $\Gamma_{Z'} / M_{Z'} \sim 2 \%$.}. We then required the signal events to be at least a $5\sigma$ fluctuation over the expected background, and in any case more than 3. This rough statistical analysis is enough to get an approximate answer to the questions we want to address. We leave a more careful analysis to the experimental collaborations ATLAS and CMS,  which have control on all the information needed to perform it in an accurate and reliable way. A more refined analysis would also be needed for a possible $Z'$ diagnostics after discovery, as studied for example in \cite{diagn}.

%
\FIGURE[t]{
\includegraphics[scale=.13]{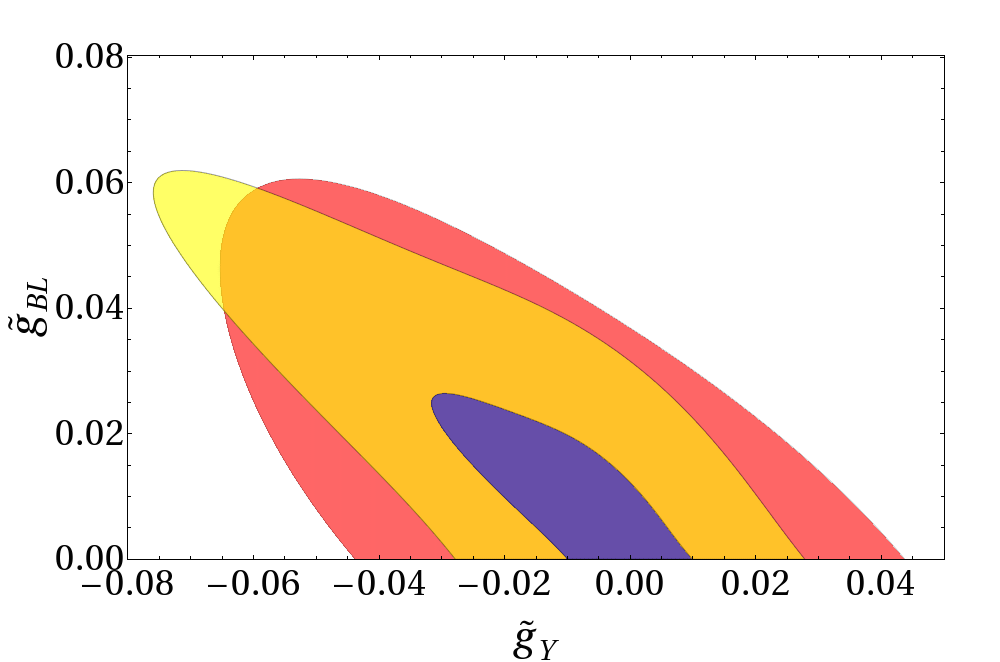}
\includegraphics[scale=.13]{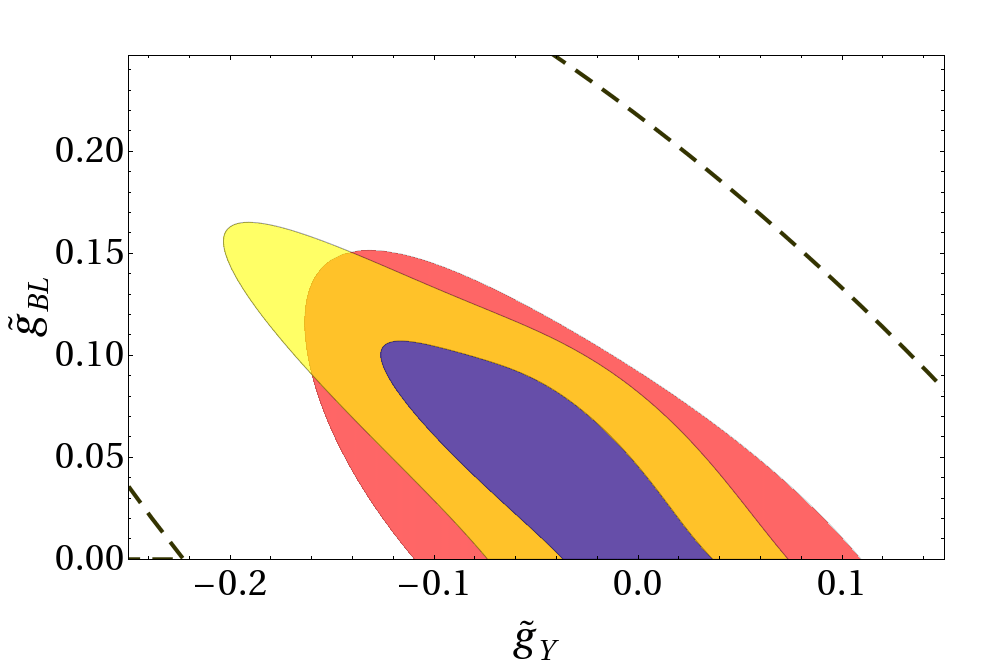}
\includegraphics[scale=.13]{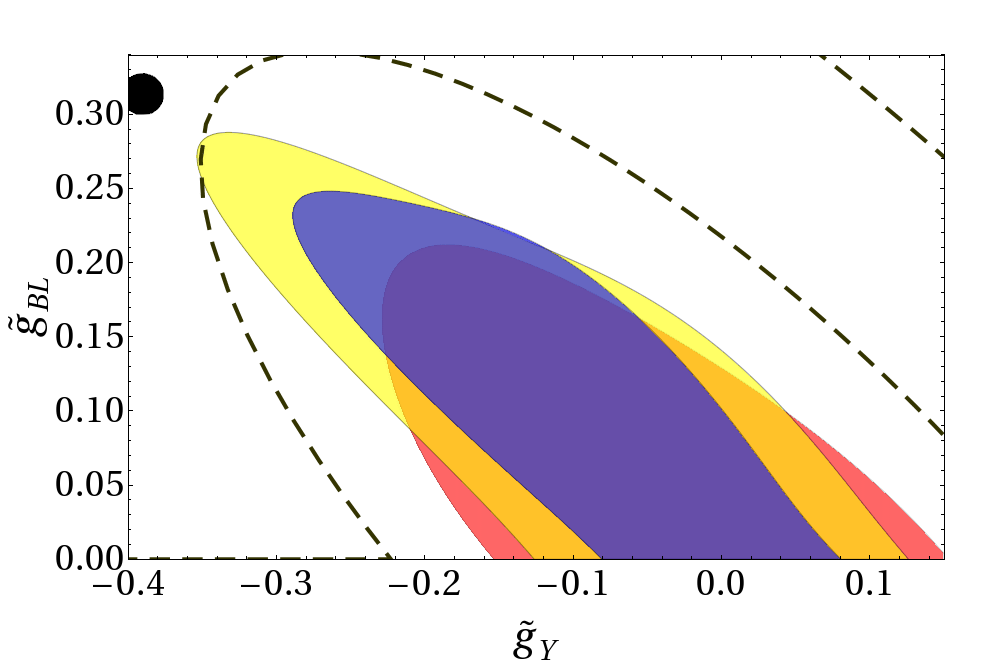}\\
\includegraphics[scale=.13]{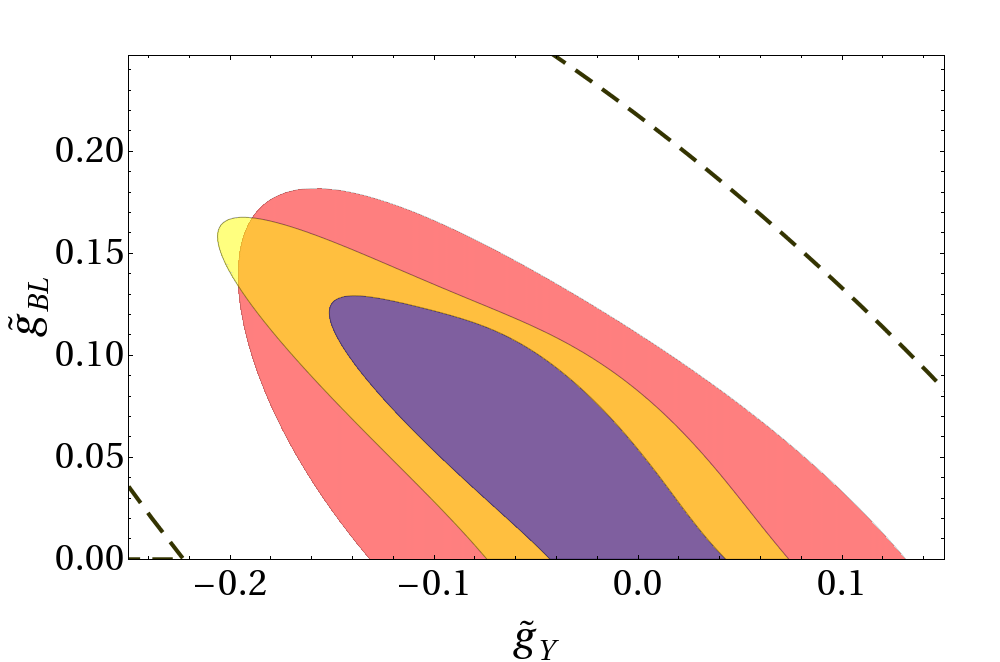}
\includegraphics[scale=.13]{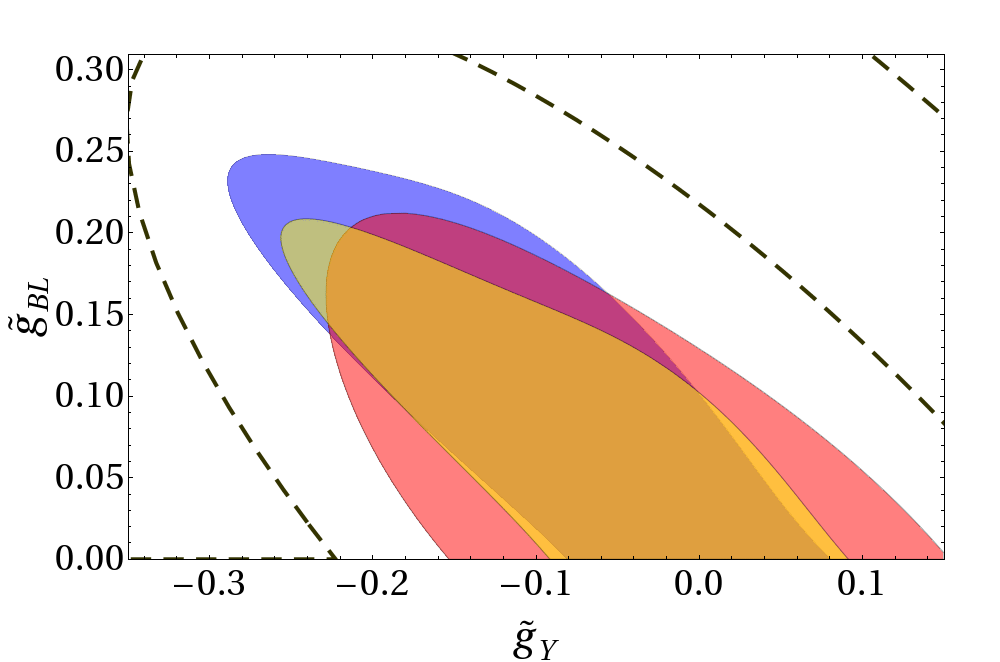}
\includegraphics[scale=.13]{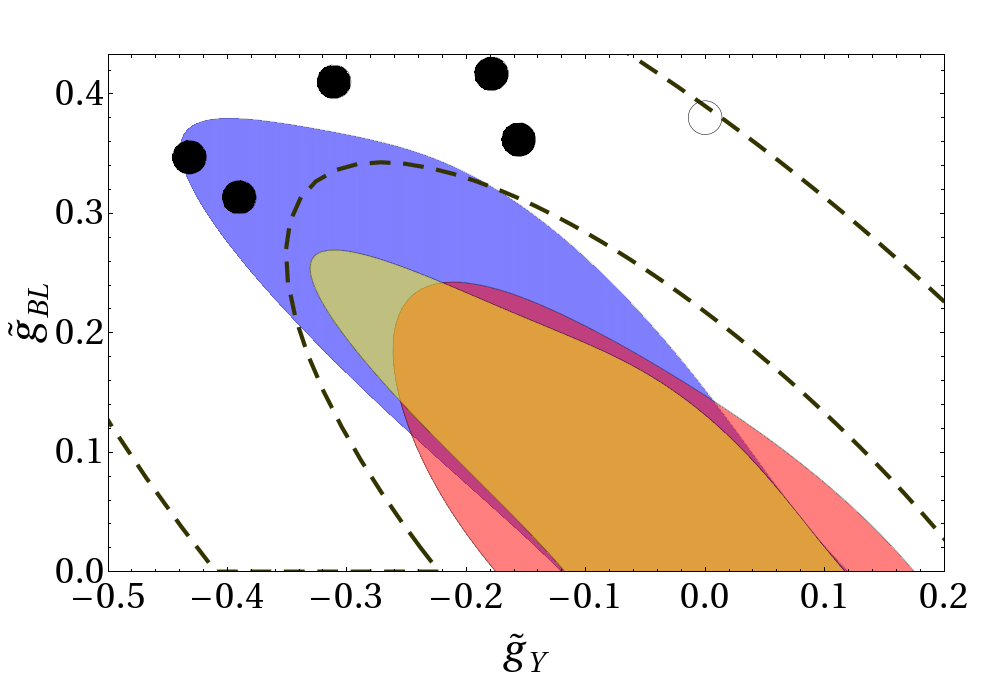}\\
\label{fig-LHC1}
\caption{The LHC 5$\sigma$ discovery potential in the $(\widetilde g_Y,\widetilde g_{BL})$ plane for $\sqrt{s}=7$~TeV. The red and blue regions are those allowed by EWPT and Tevatron bounds respectively; the yellow region is the one not within 5$\sigma$ discovery reach at the LHC. Thus the region accessible by the LHC is the one formed by points that are both in the red and blue regions but not in the yellow one. Plots in the first row refer to 50~pb$^{-1}$ of data and $M_{Z'}=200$, $500$, $700$~GeV respectively; plots in the second row are for 100~pb$^{-1}$ of data and $M_{Z'}=600$, $700$, $800$~GeV respectively.}
}
\FIGURE[t] {
\includegraphics[scale=.13]{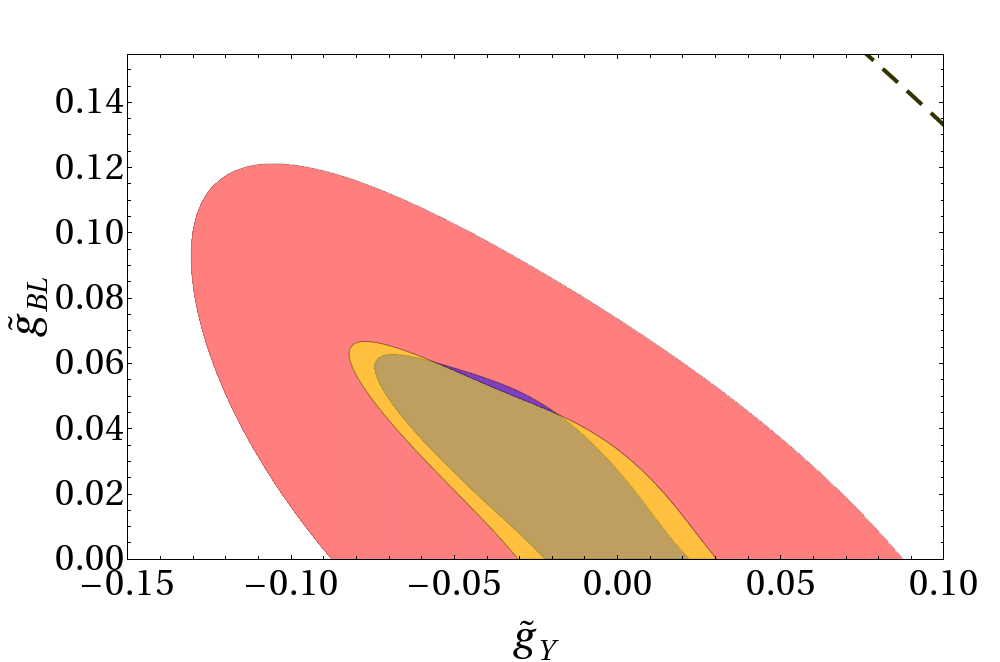}
\includegraphics[scale=.13]{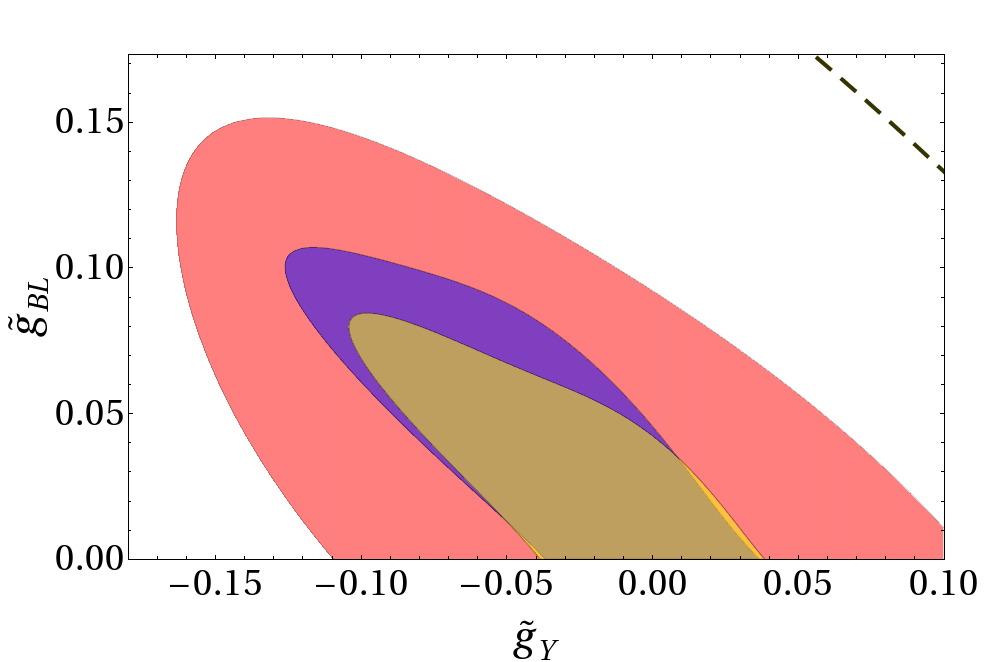}
\includegraphics[scale=.13]{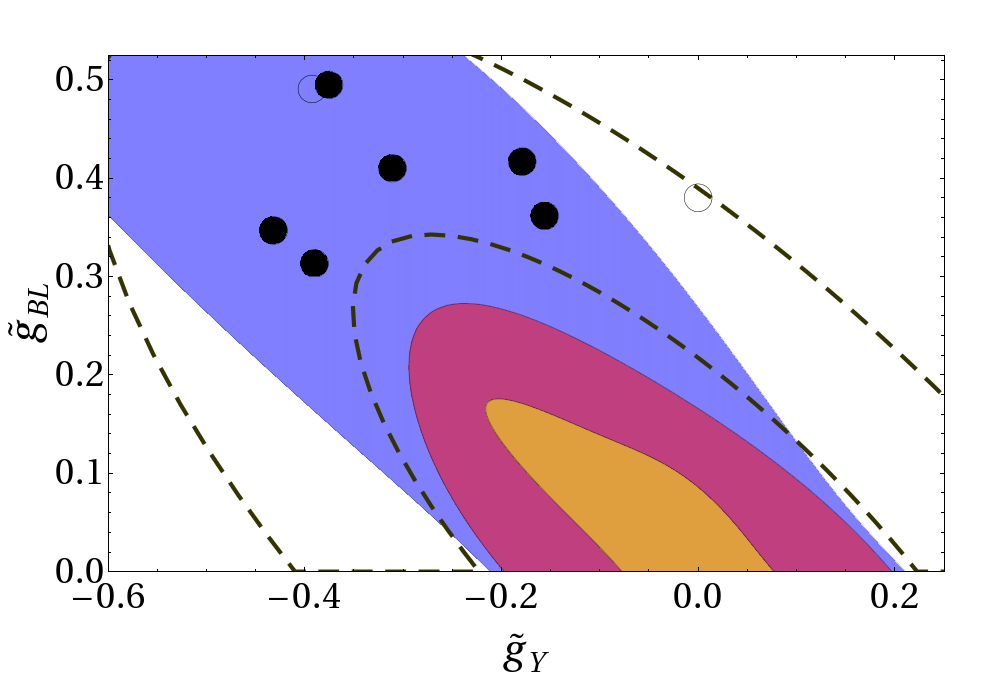}
\includegraphics[scale=.13]{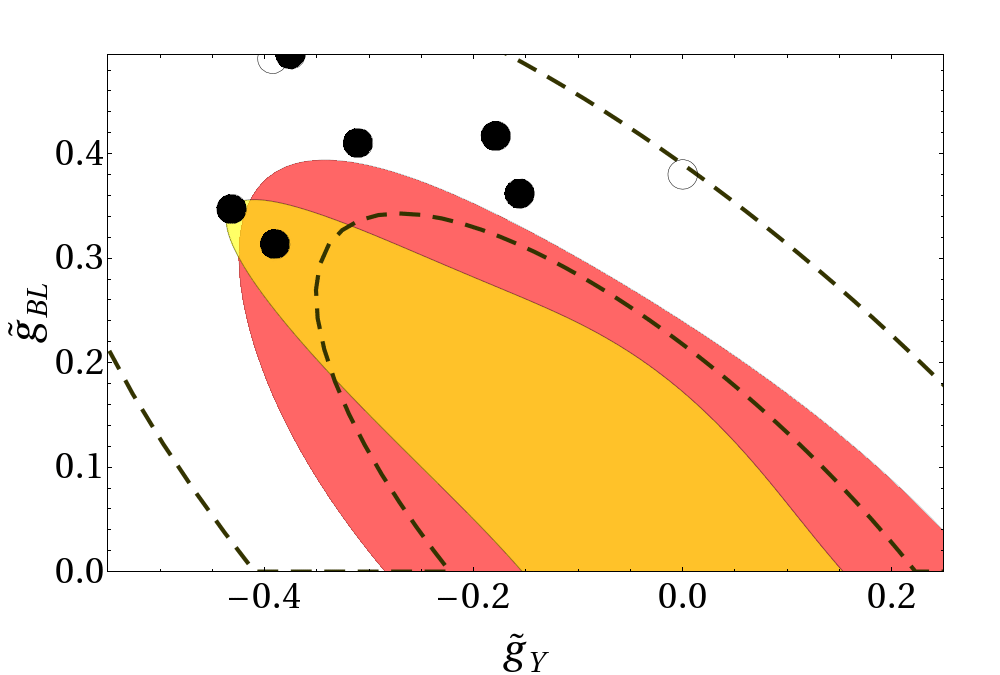}
\includegraphics[scale=.13]{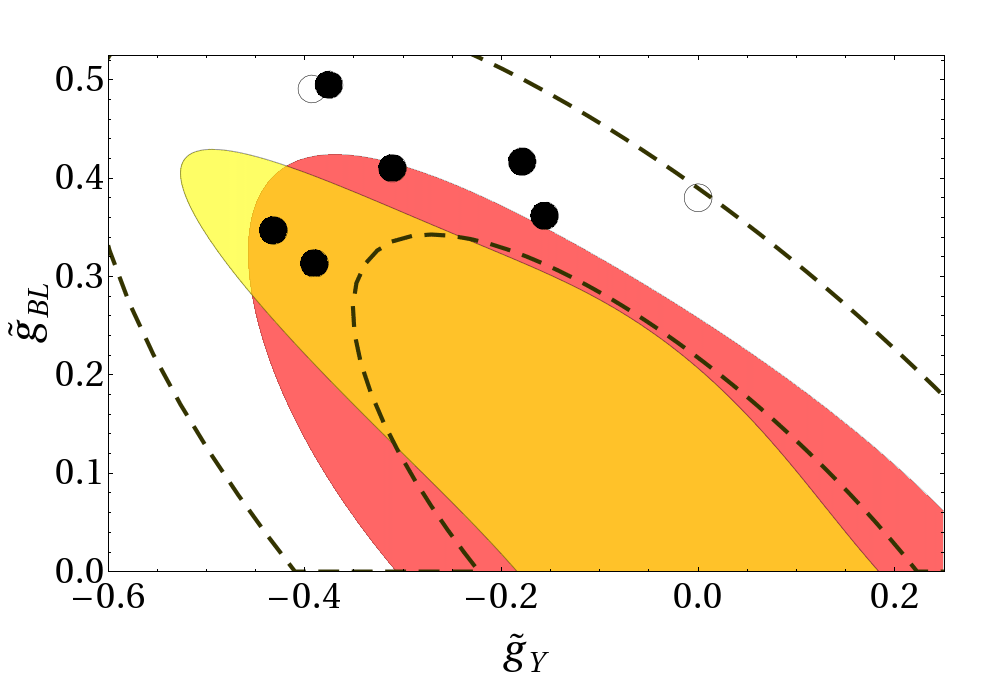}
\includegraphics[scale=.13]{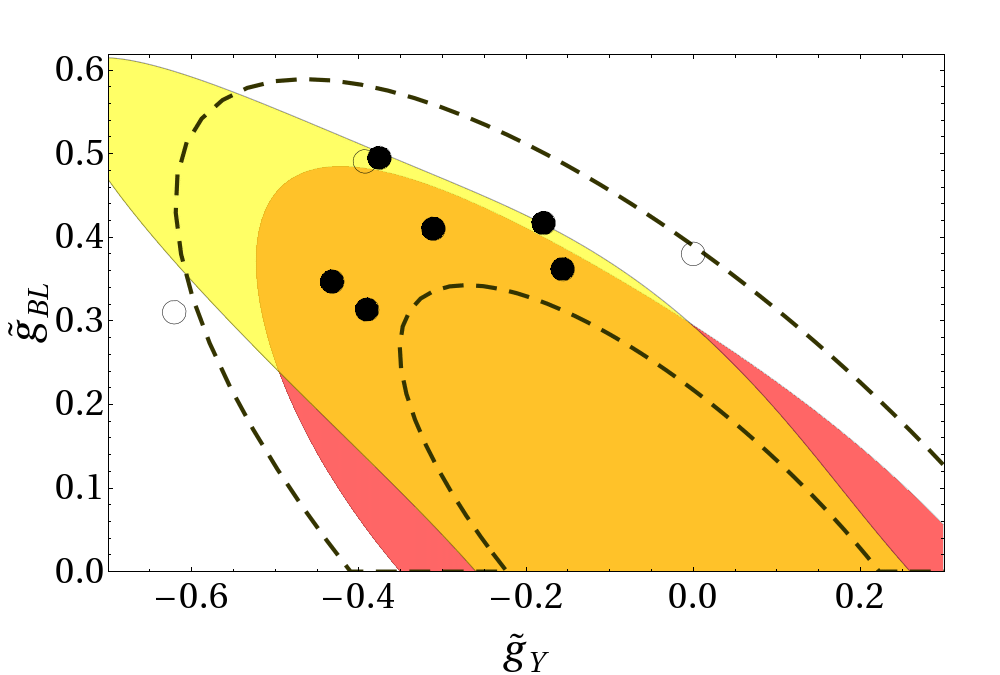}
\label{fig-LHC2}
\caption{The LHC 5$\sigma$ discovery potential in the $(\widetilde g_Y,\widetilde g_{BL})$ plane for $\sqrt{s}=10$~TeV, 200~pb$^{-1}$ of data and $M_{Z'}=400$, $500$, $900$, $1300$, $1400$, $1600$~GeV. The meaning of the colored regions is as in Fig.~\ref{fig-LHC1}. In the last three plots the Tevatron bounds are not shown because they are too weak to give useful constraints.}
}

We present sample plots in Figs.~\ref{fig-LHC1} and \ref{fig-LHC2}. Besides the regions allowed by EWPT and Tevatron data (red and blue) we plotted, for each representative value of the  $Z'$ mass, and of the LHC energy and integrated luminosity, the region not accessible to the LHC (in yellow) for a 5$\sigma$ discovery as defined above. We see that $\sqrt{s}=7$~TeV and 50~pb$^{-1}$ are not enough to discover any $Z'$ in the whole parameter space considered; in particular at low masses the low luminosity makes the LHC underperform with respect to the Tevatron, while at masses where the LHC starts having a kinematical advantage over the Tevatron, both cannot compete anymore with EWPT. With 100~pb$^{-1}$ of data at the same energy, a first non-excluded region of parameters  becomes accessible to discovery, though it is very narrow ($M_{Z'}\sim 700\pm100$~GeV and $\widetilde g_{BL}\sim 0.15\div 0.20$,
$\widetilde g_{Y}\sim -0.2 \div 0$).

Things start improving as the LHC steps up in energy and luminosity. The situation with $\sqrt{s}=10$~TeV and 200~pb$^{-1}$ of integrated luminosity  is represented in Fig.~\ref{fig-LHC2}. The region of $Z'$ masses below 400~GeV will not be accessible yet, this because the higher luminosity collected at the Tevatron is more important in such energy region. The first accessible zone in parameter space starts showing up for $M_{Z'}\sim400\div1100$~GeV, for models with couplings smaller than those preferred by GUTs, and for $M_{Z'}\sim 1200\div1500$~GeV for GUT-like couplings. For heavier $Z'$ no region is left to the LHC that is not already ruled out by EWPT. As evident from the plots, for each of the accessible $Z'$ masses, only a small portion of the $(\widetilde g_Y,\widetilde g_{BL})$ plane will be tested. Our plots refer to data collected 
by a single experiment and for a single dilepton channel,  combining the data might
help increasing the effective luminosity collected  and thus the discovery potential.

Since contour plots may require some patience to be interpreted, we make our results more manifest by plotting, in Fig.~\ref{fig-chi2}, the 5$\sigma$ LHC discovery potential
\FIGURE[t]{
\includegraphics[scale=0.2]{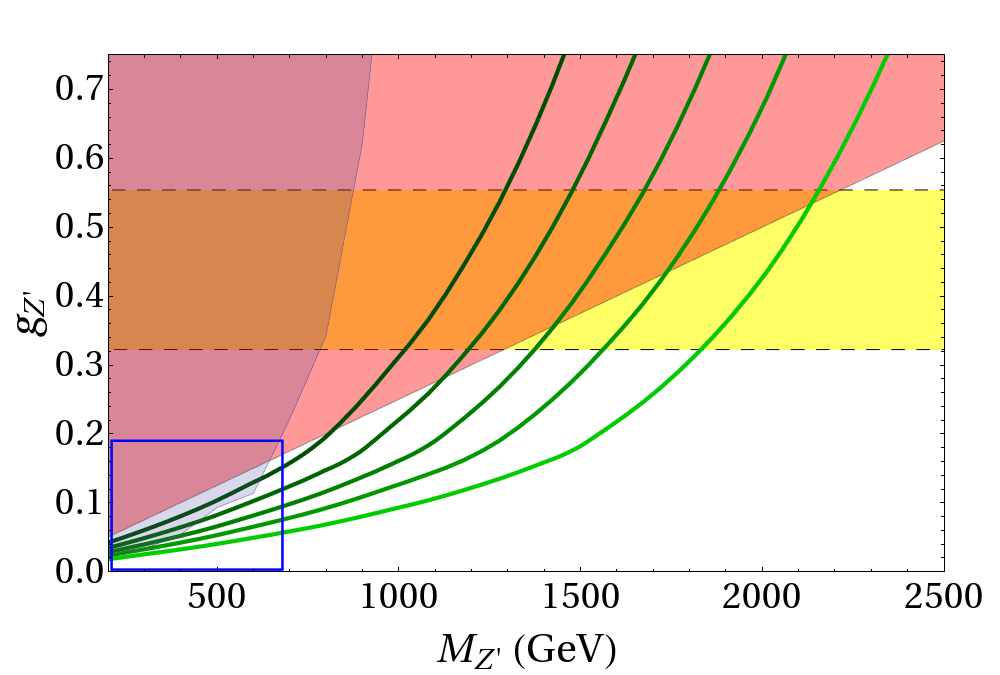} 
\includegraphics[scale=0.2]{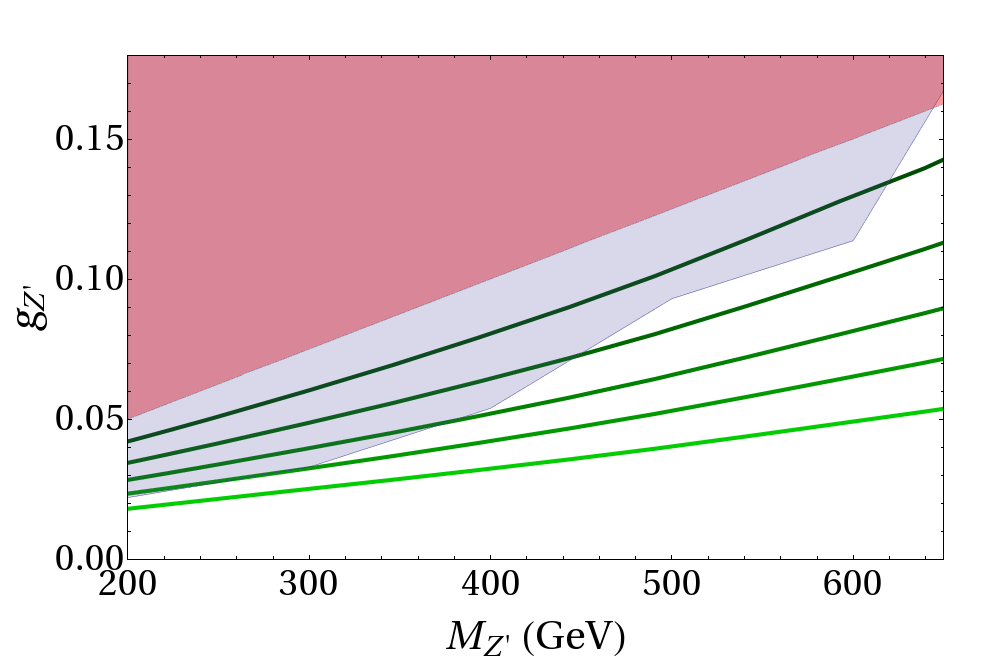} \\
\includegraphics[scale=0.2]{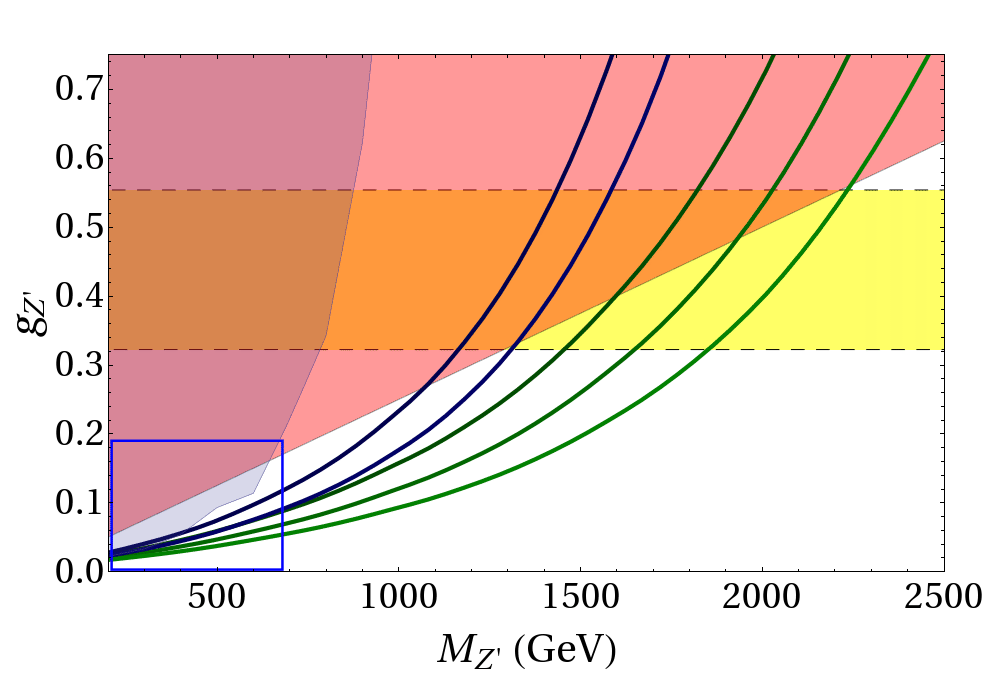} 
\includegraphics[scale=0.2]{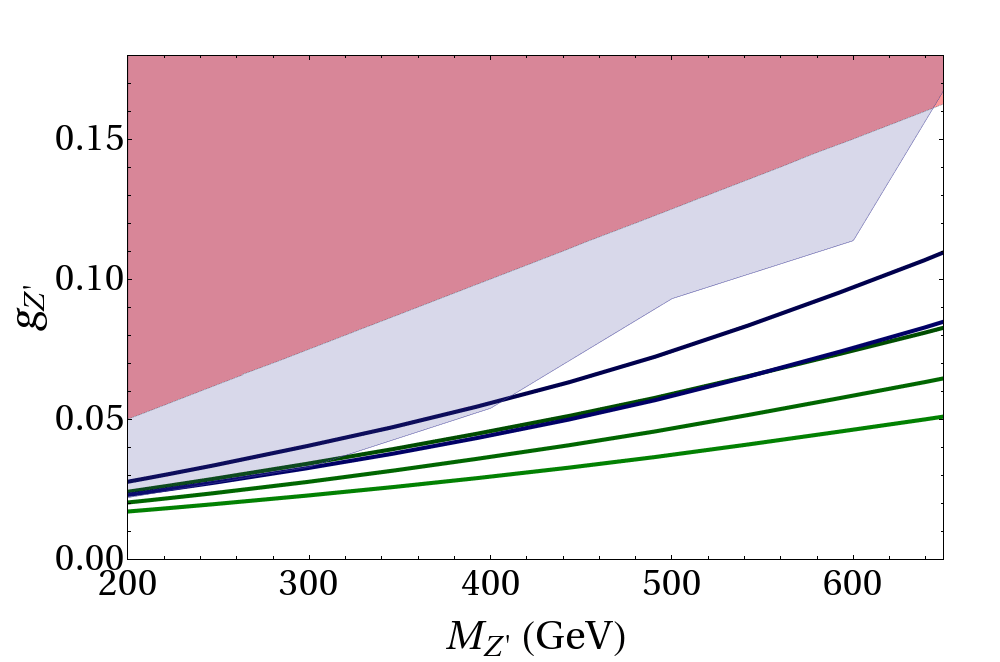} 
\label{fig-chi2}
\caption{\emph{First row.} The region of the $(M_{Z'}, g_{Z^\prime})$ plane amenable to a `5$\sigma$' discovery at the LHC, for the $Z_\chi$ model, $\sqrt{s} = 10 \, {\rm TeV}$ and some representative values of the integrated luminosity; from left to right: 50, 100, 200, 400 and 1000 pb$^{-1}$. The red and blue region and the yellow band are the same as in Fig.~\ref{fig-chi1}. The second box is a zoom on the low-mass, low-coupling region. \emph{Second row.} 95\% CL exclusion contours from the LHC after 50 and 100 pb$^{-1}$ at $\sqrt{s}=7$~TeV (blue curves) and after 50, 100 and 200~pb$^{-1}$ at $\sqrt{s}=10$~TeV (green curves).}
}
in the $(M_{Z'},g_{Z'})$ plane for the representative $\chi$ model. As in Fig.~\ref{fig-chi1}, the red and blue regions are those presently excluded by EWPT and Tevatron direct searches, respectively, and the yellow band denotes the GUT-favored region. The new curves enclose the region where a 5$\sigma$ discovery at the LHC is in principle possible, for $\sqrt{s} = 10 \, {\rm TeV}$ and some representative values of the integrated luminosity: 50, 100, 200, 400 and 1000 ${\rm pb}^{-1}$, from left to right. Notice that, in the case under consideration, the first mass region to be touched is between 600 and 800~GeV, with the region enlarging towards higher mass values with increasing luminosity. We start entering the GUT-favored region of parameters only for 100~${\rm pb}^{-1}$, and 1~${\rm fb}^{-1}$ is enough to reach mass values as high as 2~TeV, with a full coverage of the GUT-favored region of couplings.  Notice also that the access to lower mass values is also gradual, and that 1~fb$^{-1}$ is required to do better than the Tevatron at $M_{Z'} \sim 200$~GeV. We have performed a similar analysis for $\sqrt{s} = 7$~TeV: in such a case, 400~pb$^{-1}$ give approximately the same sensitivity as 200~pb$^{-1}$ at 10~TeV for $M_{Z'} < 700$~GeV, whilst the sensitivity at higher mass values rapidly becomes worse and worse, as expected: there are no doubts that it is worth raising the LHC energy as soon as it can be safely done.

Of course, if no discrepancy from the SM is found in the dilepton spectra, LHC will be 
able to improve the 95\% CL bounds on minimal $Z'$ models already after the first run(s). Indeed, as shown in Fig.~\ref{fig-chi2}, 100~pb$^{-1}$ at $\sqrt{s}=7$~TeV are already enough to top both Tevatron and EWPT bounds for all $Z'$ masses up to 1.3~TeV, while, after the first year of run, LHC might be able to rule out most of the GUT-preferred region below $\sim2$~TeV.

In summary, our study shows how strong the r\^ole played by the existing experimental bounds can be in limiting the access to new physics in the early LHC phase. Even for the `easy' $Z'$ models usually considered, the energy and luminosity required to overcome existing bounds can delay the possibility of discovering new physics by a non-negligible amount. The importance of reaching higher energies and luminosities is clear, as it is the importance of combining data from different detectors and channels already in this early phase.

Our analysis also shows that different regions in parameter space will become available for discovery at different times, depending on the energies and luminosity reached. Hence it would be sensible to switch to general parameterizations such as the one described in the present work, which are not as restrictive as those commonly used at the moment. The latter may focus the attention on the `wrong' regions of parameter space that are already ruled out by current data, instead of those with the greatest potential to host accessible new physics. 

%
%
\acknowledgments
We thank H.~Burkhardt, R.~Contino, S.~Forte, L.~Ib\'a\~nez, M.~Pierini, G.~Ridolfi, A.~Strumia, E.~Torassa and A.~Uranga for discussions.
This work was partially supported by the Fondazione Cariparo Excellence Grant {\em String-derived supergravities with branes and fluxes and their phenomenological implications}.
%


%
\end{document}